\documentclass[reprint, amsmath,amssymb, unsortedaddress, aps, pre]{revtex4-2}

\usepackage{svg}
\usepackage[utf8]{inputenc}
\usepackage{subfigure}
\usepackage{comment}
\usepackage{mathptmx}
\usepackage{amsmath, bm}
\usepackage{multirow}
\usepackage{hyperref}

\usepackage{graphicx}
\usepackage{dcolumn}
\usepackage{bm}

\begin{document}


\title{Radiation Forces and Torques on Janus Cylinders}


\author{Mohd. Meraj Khan}
 \email{am19d041@smail.iitm.ac.in}
\affiliation{%
 Department of Applied Mechanics and Biomedical Engineering,
 Indian Institute of Technology Madras, Chennai, India
}%

\author{Sumesh P. Thampi}
 \email{sumesh@iitm.ac.in}
\affiliation{
 Department of Chemical Engineering, Indian Institute of Technology Madras, Chennai, India
}%

\author{Anubhab Roy}%
 \email{Corresponding author: anubhab@iitm.ac.in}
\affiliation{%
 Department of Applied Mechanics and Biomedical Engineering,
 Indian Institute of Technology Madras, Chennai, India
}%


\begin{abstract}

We investigate radiation-induced drag, lift, and torque on circular Janus cylinders under transverse-magnetic plane-wave illumination, considering metallo-dielectric and purely dielectric configurations. The lattice Boltzmann method (LBM) is employed with absorption neglected, isolating scattering as the sole momentum-transfer mechanism. For metallo-dielectric Janus cylinders, analytical expressions for radiation force and torque are derived and used to validate the LBM, showing excellent agreement across a wide range of dielectric constants and interface orientations. For dielectric Janus cylinders, material inhomogeneity induces asymmetric scattering giving rise to nonzero lift and torque under plane-wave illumination, with non-monotonic dependence on interface orientation and dielectric contrast. Two mechanisms govern the observed variations: resonance-driven energy amplification and scattered field redistribution. The computed force and torque maps serve as design diagrams for predicting the optomechanical response. Coupling these with viscous dynamics at low Reynolds number reveals diverse particle trajectories, including curved paths during reorientation and nearly straight motion once torque-free equilibria are reached. The system is externally actuated and results represent scattering-dominated dynamics under idealized conditions, providing physical insight into optomechanical responses of Janus particles with implications for trajectory shaping in optofluidic systems.

\end{abstract}

\maketitle



\section{\label{sec:introduction}Introduction}

When an electromagnetic wave encounters a particle with properties differing from the surrounding medium, the particle redirects energy from the incident wave in all directions, resulting in scattered radiation \cite{kerker1969scattering}. Light carries both energy and momentum \cite{jackson1999classical, griffiths2005introduction}, and the transfer of momentum during scattering and absorption gives rise to radiation force and torque on the particle. These forces and torques follow from momentum conservation and depend on the scattering and absorption characteristics of the particle as well as the properties of the incident field \cite{bohren1998absorption, mishchenko2002scattering}.

Radiation forces have found wide-ranging applications, particularly since the development of lasers. Optical tweezers enable precise three-dimensional manipulation of microscopic objects and are widely used in biophysics, cell biology, and single-molecule studies \cite{ashkin1977observation, ashkin1987optical, ashkin1989optical, ashkin1992forces, bustamante2021optical}. Optical manipulation also plays an important role in nanotechnology, for example in nanoparticle assembly and nanofabrication \cite{zemanek2019perspective, chai2022directed, wang2016photodynamic}. Plasmonic nanoagents further extend these capabilities to biosensing, drug delivery, photothermal therapy, and bioimaging \cite{huergo2022plasmonic, wu2022light}. In all these applications, the magnitude and directionality of radiation forces are key to functionality.

Optical forces and torque originate from the transfer of linear and angular momentum from electromagnetic waves to matter \cite{hulst1981light, bohren1998absorption, mishchenko2002scattering}. For isotropic and homogeneous particles, symmetry restricts the radiation response to a drag force aligned with the incident wave, with no lift or torque generated under plane-wave illumination \cite{hulst1981light, bohren1998absorption}. Anisotropic or inhomogeneous structures break these symmetries and enable asymmetric scattering or absorption, giving rise to lateral forces and radiation-induced torque even for stationary particles under plane-wave illumination \cite{sukhov2017non, tanaka2020plasmonic, buzas2012light}. In contrast, additional lateral forces and torques may also arise in moving or rotating particles due to dynamical effects, which are not considered in the present study \cite{li2025optical}.

All real materials exhibit some degree of optical absorption, which can be described through a complex refractive index $\tilde{n} = n + i\kappa$, where the real part governs wave propagation and the imaginary part accounts for attenuation and energy dissipation \cite{born1999principles, bohren1998absorption}. While scattering redistributes electromagnetic momentum, absorption leads to irreversible conversion of electromagnetic energy into heat \cite{bohren1998absorption}. In metallic nanostructures, these effects are strongly enhanced by plasmonic resonances, where collective oscillations of conduction electrons produce large local field enhancements and significantly increase both scattering and absorption cross sections \cite{maier2007plasmonics, bohren1998absorption, kreibig2013optical}. These plasmonic effects, combined with structural asymmetry and material contrast, enable controlled redistribution of scattered momentum, giving rise to lateral and pulling forces as well as radiation-induced rotation \cite{sukhov2017non, yifat2018reactive, tanaka2020plasmonic}. In many experimental realizations, especially in plasmonic and thermoplasmonic systems, absorption plays a dominant role where asymmetric heating leads to self-thermophoretic motion and optically driven flows \cite{baffou2013thermo, govorov2007, jiang2010active}, and such mechanisms have been used to realize light-driven nanomotors and optically propelled particles capable of controlled translation and reorientation under illumination \cite{liu2010light, liu2016self, wu2022light, gonzalezcolsa2022nanojet, serrera20253}. A prominent class of systems combining geometric and material asymmetry is provided by Janus particles, which consist of two regions with distinct optical properties and exhibit rich optomechanical behavior under electromagnetic excitation, including thermoplasmonic propulsion in hybrid architectures \cite{gonzalezcolsa2022nanojet, serrera20253}.

The structural and material asymmetry of Janus particles leads to anisotropic scattering, enabling lateral forces and radiation-induced torque even under plane-wave illumination \cite{tanaka2020plasmonic, buzas2012light, rodriguez2015lateral}, and these properties have been explored across a variety of configurations including plasmonic and dielectric systems, with applications in optical trapping, propulsion, and directed motion \cite{li2017plasmonic, huergo2022plasmonic, zhang2017janus, soto2021reversible, koya2023resonant, xiao2018review}. Previous studies have employed numerical approaches such as the discrete dipole approximation (DDA) for dielectric Janus spheres \cite{simpson2011application} and ray-optics models for metallo-dielectric Janus particles \cite{liu2015ray}. However, in many cases the optomechanical response arises from a combination of scattering and absorption, making it difficult to isolate the contribution of scattering-induced momentum transfer \cite{baffou2013thermo, govorov2007, gonzalezcolsa2022nanojet}. As a result, a systematic full-wave analysis of radiation forces and torque in composite Janus structures remains limited, particularly in the intermediate size regime ($a/\lambda \sim 1$) where multiple scattering modes contribute comparably, producing strong interference effects and nontrivial redistribution of electromagnetic momentum that cannot be captured by ray-optics or dipole approximations \cite{hulst1981light, bohren1998absorption}. In this regime, the interface orientation angle $\phi_0$ plays a central role, since even small rotations can substantially alter the symmetry of the scattered field and hence the direction and magnitude of the resulting forces and torque, requiring full-wave solutions of Maxwell's equations for accurate prediction \cite{khan2024electromagnetic}. Among the available numerical techniques --- including finite-difference time-domain method \cite{schneider2010understanding}, finite element method \cite{jin2015finite}, DDA \cite{simpson2011application}, and T-matrix methods \cite{mishchenko2002scattering} --- the lattice Boltzmann method (LBM) provides an efficient and flexible framework for the composite geometries considered here \cite{khan2024electromagnetic, khan2026lbm_em_scattering}.

In this work, we employ LBM \cite{khan2024electromagnetic, khan2026lbm_em_scattering} to investigate radiation forces and torque on metallo-dielectric and dielectric Janus cylinders under plane-wave illumination. The cylindrical geometry is adopted because it admits analytical solutions against which the numerical method can be rigorously benchmarked \cite{hurd1975diffraction}, and because it isolates the essential physics of material-interface asymmetry in a two-dimensional setting that is directly relevant to nanowire and nanorod systems \cite{pauzauskie2006optical}. Transverse-magnetic (TM$^z$) polarization is considered throughout, as it couples strongly to the in-plane material contrast and provides the richest optomechanical response for the geometries studied \cite{hulst1981light, bohren1998absorption}; the transverse-electric case follows analogously and is left for future work. Extending our previously validated LBM framework for homogeneous scatterers \cite{khan2024electromagnetic, khan2026lbm_em_scattering} to composite geometries, we benchmark the method against analytical solutions for metallo-dielectric Janus cylinders. By isolating scattering as the sole momentum-transfer mechanism \cite{bohren1998absorption, mishchenko2002scattering}, we systematically examine the roles of geometric asymmetry and dielectric contrast in governing radiation drag, lift, and torque. The computed force and torque maps — showing the optomechanical response as functions of interface orientation angle and material contrast — serve as design diagrams that can guide the selection of material parameters and particle orientations in experimental realizations. Building on these maps, we further analyze the trajectories that free particles would follow under illumination, connecting the local force and torque distributions to emergent translational and rotational dynamics. We emphasize that the system studied here represents an externally actuated particle rather than an active one in the strict sense: the motion is driven entirely by radiation forces and torque imposed by an external electromagnetic field, without any local energy transduction within the particle itself \cite{marchetti2013hydrodynamics}. While factors such as absorption, beam intensity, and fabrication constraints are not explicitly considered, the present results offer a controlled framework for understanding scattering-driven momentum transfer in asymmetric particles \cite{bohren1998absorption, mishchenko2002scattering}, with direct implications for the design of externally actuated, optically steerable nanostructures.

Throughout, forces, torques, positions, and time are reported in non-dimensional form, with the relevant scales defined in Section~\ref{sec:theory} and Section~\ref{sec: results and discussions} respectively.

The structure of the paper is as follows. Section~\ref{sec:theory} presents the theoretical formulation. Section~\ref{sec:LBM} describes the numerical implementation. Section~\ref{sec: results and discussions} presents validation, analytical results, and parametric studies, including trajectory analysis. Section~\ref{sec: Conclusion} summarizes the main findings.


\section{Scattering Model and Radiation Force Formulation} \label{sec:theory}

\begin{figure}[!htb]
    {\includegraphics[]{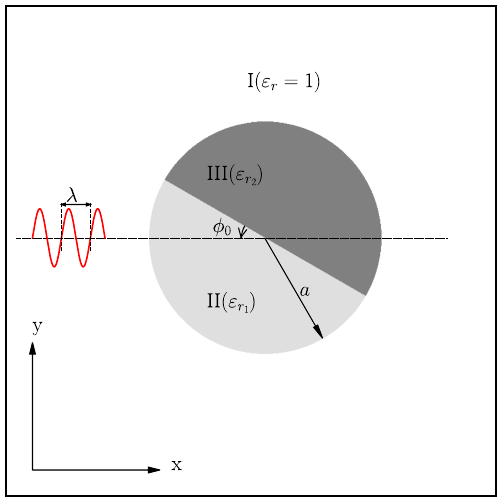}}
    \caption{Schematic of the computational domain for a circular Janus cylinder of radius $a$, divided into two halves with different dielectric properties. The cylinder is placed in vacuum and illuminated by a plane wave of wavelength $\lambda$ incident at angle $\phi_0$ relative to the Janus interface.}
    \label{fig:schematic_LBM}
\end{figure}

We consider a circular Janus cylinder of radius $a$, infinitely long along the $z$-axis, as shown in Fig.~\ref{fig:schematic_LBM}. The cylinder is embedded in vacuum (permittivity $\varepsilon_0$ and permeability $\mu_0$) and consists of two regions with distinct dielectric properties. Two configurations are studied: (i) a metallo-dielectric Janus cylinder, where one half (region III) is a perfect electric conductor (PEC) and the other (region II) has dielectric constant $\varepsilon_{r_1}$, and (ii) a dielectric Janus cylinder, where the two halves have dielectric constants $\varepsilon_{r_1}$ (region II) and $\varepsilon_{r_2}$ (region III), respectively. In both cases, the surrounding vacuum constitutes region I.

The cylinder is illuminated by a monochromatic plane wave of wavelength $\lambda$ with TM$^z$ polarization, incident at an angle $\phi_0$ relative to the Janus interface. The incident electric field is polarized along the $z$-axis and can be expressed as
\begin{equation} \label{eq: incident field}
    E_z^I (r, \phi) = \sum_{m=0}^{\infty} \alpha_m (-j)^m J_m(k_0 r) \cos m(\phi - \phi_0),
\end{equation}
where $J_m$ is the Bessel function of the first kind, $k_0 = 2\pi/\lambda$ is the free-space wavenumber, and $\alpha_m = 1$ for $m = 0$ and $\alpha_m = 2$ for $m \geq 1$.

The scattered electric field in region I is given by
\begin{equation} \label{eq: scattered field}
    E_z^S (r, \phi) = \sum_{m=0}^{\infty} \left[ B_m \cos(m\phi) + D_m \sin(m\phi) \right] H_m^{(1)}(k_0 r),
\end{equation}
where $H_m^{(1)}(k_0r)$ is the Hankel function of the first kind of order $m$. The coefficients $B_m$ and $D_m$ are obtained by enforcing boundary conditions at the interface. Time dependence $e^{-j\omega t}$ is omitted throughout.

\subsection{Far-field formulation} \label{sec:SW}

To characterize scattering, we evaluate the far-field intensity derived from the asymptotic behavior of the scattered field \cite{hulst1981light, bohren1998absorption}. In the limit $r \gg a$, where $r$ is the radial distance from the cylinder axis, the scattered field in two dimensions decays as $e^{-jk_0r}/\sqrt{r}$ \cite{hulst1981light, bohren1998absorption}. This enables the definition of an angular scattering intensity as a function of the observation angle $\phi$ \cite{hulst1981light, bohren1998absorption}:
\begin{equation} \label{eq: scattering intensity}
    |S(\phi)|^2 = k_0 r \frac{|\mathbf{E}^S(\phi)|^2}{|\mathbf{E}^I|^2},
\end{equation}
where $\mathbf{E}^S(\phi)$ is the scattered electric field evaluated at the far-field observation point at distance $r$ in the direction $\phi$, and $\mathbf{E}^I$ is the amplitude of the incident electric field evaluated in the vicinity of the scatterer. This quantity is independent of the observation distance $r$.

For analytical reference, the far-field form is obtained using the large-argument approximation of the Hankel function of the first kind $H_m^{(1)}(k_0 r)$, valid for $k_0 r \to \infty$ \cite{hulst1981light, bohren1998absorption}:
\begin{equation*}
    H_m^{(1)} (k_0 r) \sim \sqrt{\frac{2}{\pi k_0 r}} \, e^{j \left(k_0 r - \frac{m\pi}{2} - \frac{\pi}{4} \right)},
\end{equation*}
where $m$ is the mode index. Substituting this approximation into Eq.~\eqref{eq: scattered field} yields the analytical far-field scattered electric field in terms of the coefficients $B_m$ and $D_m$, which are determined by enforcing boundary conditions at the cylinder surface, as detailed in Section~\ref{sec:MDJC}. Since far-field quantities cannot be directly extracted from the finite computational domain, a near-to-far-field transformation is employed to extrapolate the LBM near-field data to the far field \cite{khan2026lbm_em_scattering, schneider2010understanding}. The resulting far-field scattering intensity serves as the primary quantity for validating the LBM against analytical solutions in Section~\ref{sec:MDJC}.

\subsection{Radiation Force and Torque via Maxwell Stress Tensor} 
\label{sec:stress_tensor}

The radiation force and torque per unit length, averaged over one time period, exerted on the cylinder are determined by integrating the time-averaged Maxwell stress tensor over its surface \cite{mishchenko2002scattering, bohren1998absorption}. In vacuum, the time-averaged Maxwell stress tensor is defined as \cite{mishchenko2002scattering}:
\begin{equation}\label{eq: stress tensor average}
    \langle \mathbb{T} \rangle = \frac{1}{2} \Re \left[ \varepsilon_0 
    {\bf E} {\bf E}^* + \mu_0 {\bf H} {\bf H}^* - \frac{1}{2} \left( 
    \varepsilon_0 |{\bf E}|^2 + \mu_0 |{\bf H}|^2 \right) \mathbb{I} 
    \right],
\end{equation}
where ${\bf E}$ and ${\bf H}$ represent the total (scattered + incident) electric and magnetic fields, respectively, in the frequency domain, and ${\bf E}^*$ and ${\bf H}^*$ are their complex conjugates. Here, $\varepsilon_0$ and $\mu_0$ denote the permittivity and permeability of the surrounding medium (free space in this case), $\mathbb{I}$ is the identity tensor, and $\Re$ denotes the real part. The notation $\langle \cdot \rangle$ represents time-averaging over one period.

The time-averaged force per unit length acting on the cylinder is computed by integrating the Maxwell stress tensor over its circumference 
\cite{mishchenko2002scattering, hulst1981light}:
\begin{equation} \label{eq: force}
    \langle {\bf F} \rangle = \oint \langle \mathbb{T} \rangle \cdot 
    \hat{{\bf n}} \, dl,
\end{equation}
where $\hat{\bf n}$ is the outward unit normal to the cylinder surface and $dl$ is the line element along the circumference. Similarly, the time-averaged torque per unit length about the cylinder axis is given by \cite{mishchenko2002scattering, bohren1998absorption}:
\begin{equation} \label{eq: torque}
    \langle {\bf M} \rangle = \oint {\bf r} \times ( \langle \mathbb{T} 
    \rangle \cdot \hat{{\bf n}} ) \, dl,
\end{equation}
where ${\bf r}$ is the position vector relative to the cylinder axis.

The total electric and magnetic fields at the cylinder surface are required for radiation force and torque calculations. The total electric field is obtained as the sum of the incident and scattered electric fields using Eqs.~\eqref{eq: incident field} and \eqref{eq: scattered field}. The total magnetic field is then determined via Maxwell's curl equation:
\begin{equation*}
{\boldsymbol \nabla} \times {\bf E} = - j \omega \mu_0 {\bf H},
\end{equation*}
where $\omega$ is the angular frequency of the incident wave. For TM$^z$ polarization, $E_r (r, \phi) = E_{\phi} (r, \phi) = H_z (r, \phi) = 0$, and the remaining components of the total electric and magnetic fields at $r = a$ are as follows 
\cite{hulst1981light, bohren1998absorption}:
\begin{widetext}
    \begin{subequations}\label{eq:total_field}
    \begin{align}
        E_z (\phi) = \sum_{m = 0}^{\infty} \alpha_m (-j)^m J_m (k_0 a) \cos\left[ m (\phi - \phi_0) \right] + \left\{ B_m \cos(m \phi) + D_m \sin(m \phi) \right\} H_m^{(1)} (k_0 a), \\
        H_r (\phi) = \frac{-j}{\omega \mu_0 a} \sum_{m = 0}^{\infty} m \alpha_m (-j)^m J_m (k_0 a) \sin\left[ m (\phi - \phi_0) \right]  + m \left\{ B_m \sin(m \phi) - D_m \cos(m \phi) \right\} 
        H_m^{(1)} (k_0 a), \\
        H_{\phi} (\phi) = \frac{-j k_0}{\omega \mu_0} \sum_{m = 0}^{\infty} \alpha_m (-j)^m J_m' (k_0 a) \cos\left[ m (\phi - \phi_0) \right]  + \left\{ B_m \cos(m \phi) + D_m \sin(m \phi) \right\} 
        H_m^{(1)'} (k_0 a),
    \end{align}
    \end{subequations}
\end{widetext}
where primes denote differentiation with respect to the argument. By substituting the total electric and magnetic fields into Eqs.~\eqref{eq: stress tensor average}, \eqref{eq: force}, and \eqref{eq: torque}, and transforming from polar to Cartesian coordinates, the radiation drag, lift, and torque are obtained as:
\begin{widetext}
    \begin{align}
        \langle F_x \rangle &= \int_0^{2 \pi} \left[ \left\{ 
        - \frac{1}{4} \varepsilon_0 E_z E_z^* + \frac{1}{4} \mu_0 
        \left( H_r H_r^* - H_{\phi} H_{\phi}^* \right) \right\} 
        \left( - \cos{\phi_0} \cos{\phi} - \sin{\phi_0} \sin{\phi} 
        \right) \right. \nonumber \\
        &\qquad \qquad \left. + \frac{1}{2} \Re \left( \mu_0 H_{\phi} 
        H_r^* \right) \left( - \sin{\phi_0} \cos{\phi} + \cos{\phi_0} 
        \sin{\phi} \right) \right] a \, d\phi, \label{eq: drag} \\
        \langle F_y \rangle &= \int_0^{2 \pi} \left[ \left\{ 
        - \frac{1}{4} \varepsilon_0 E_z E_z^* + \frac{1}{4} \mu_0 
        \left( H_r H_r^* - H_{\phi} H_{\phi}^* \right) \right\} 
        \left( \sin{\phi_0} \cos{\phi} - \cos{\phi_0} \sin{\phi} 
        \right) \right. \nonumber \\
        &\qquad \qquad \left. + \frac{1}{2} \Re \left( \mu_0 H_{\phi} 
        H_r^* \right) \left( - \cos{\phi_0} \cos{\phi} - \sin{\phi_0} 
        \sin{\phi} \right) \right] a \, d\phi \label{eq: lift} \\
        \langle M_z \rangle &= \int_0^{2 \pi} \frac{1}{2} \mu_0 a^2 
        \Re ( H_{\phi} H_r^* ) \, d \phi, \label{eq: Torque}
    \end{align}
\end{widetext}
where, $\langle F_x \rangle$ represents the time-averaged radiation drag, $\langle F_y \rangle$ the radiation lift, and $\langle M_z \rangle$ the radiation torque, with all quantities expressed per unit length. The non-dimensional forms of these quantities are defined as
\begin{equation*}
    \tilde{F}_x = \frac{\langle F_x \rangle}{F_0}, \quad
    \tilde{F}_y = \frac{\langle F_y \rangle}{F_0}, \quad
    \tilde{M}_z = \frac{\langle M_z \rangle}{M_0},
\end{equation*}
with reference scales $F_0 = \pi a \varepsilon_0 E_0^2 / 4$ and $M_0 = 2 F_0 a$, where $E_0$ is the amplitude of the incident electric field, set to unity in the non-dimensional formulation of Eq.~\eqref{eq: incident field}.


\section{Lattice Boltzmann Method} \label{sec:LBM}

We employ LBM to solve Maxwell’s equations on a D3Q7 lattice \cite{hauser2017stable}, where $\mathbf{e}_i({\bf r}, t)$ and $\mathbf{h}_i({\bf r}, t)$ denote the distribution functions associated with the electric and magnetic fields, respectively, and the subscript $i$ corresponds to the discrete lattice velocity directions. These distribution functions evolve according to the lattice Boltzmann equations:
\begin{subequations} \label{eq: LBGK equation}
\begin{gather}
    \mathbf{e}_{i}({\bf r} + {\bf c}_i\Delta t, t + \Delta t) = 2 \mathbf{e}_{i}^{eq}({\bf r} ,t) - \mathbf{e}_{i}({\bf r} ,t), \\
    \mathbf{h}_{i}({\bf r} + {\bf c}_i\Delta t, t + \Delta t) = 2 \mathbf{h}_{i}^{eq}({\bf r} ,t) - \mathbf{h}_{i}({\bf r} ,t),
\end{gather}
\end{subequations}
where $\Delta t$ is the time step and ${\bf c}_i$ are the lattice velocities. The equilibrium distribution functions $\mathbf{e}_i^{eq}$ and $\mathbf{h}_i^{eq}$ are defined as \cite{hauser2017stable}
\begin{subequations} \label{eq: equilibrium distribution functions}
\begin{gather}
    \mathbf{e}_i^{eq} ({\bf r} ,t) = 
    \begin{cases}
      \frac{1}{6}\Big(\pmb{ \mathcal{E}} ({\bf r} ,t) -   {\bf c}_i \times \pmb{ \mathcal{H}} ({\bf r} ,t) \Big) & \text{if $i \neq 0$}\\
      (\varepsilon_{r} - 1 ) \pmb{ \mathcal{E}} ({\bf r} ,t) & \text{if $i = 0$}
    \end{cases},  \\
    \mathbf{h}_i^{eq} ({\bf r} ,t) = 
    \begin{cases}
      \frac{1}{6}\Big(\pmb{ \mathcal{H}} ({\bf r} ,t) +  {\bf c}_i \times \pmb{ \mathcal{E}} ({\bf r} ,t) \Big) & \text{if $i \neq 0$}\\
      (\mu_{r} - 1 ) \pmb{ \mathcal{H}} ({\bf r} ,t) & \text{if $i = 0$}
    \end{cases},  
\end{gather}
\end{subequations}
where $\varepsilon_r$ and $\mu_r$ denote the relative permittivity and permeability respectively. The macroscopic electric and magnetic fields, $\pmb{\mathcal{E}}$ and $\pmb{\mathcal{H}}$, are reconstructed from the moments of their respective distribution functions \cite{hauser2017stable}. Here, $\pmb{\mathcal{E}}(\mathbf{r},t)$ and $\pmb{\mathcal{H}}(\mathbf{r},t)$ denote the time-domain electric and magnetic fields; the corresponding frequency-domain fields $\mathbf{E}$ and $\mathbf{H}$, used in the far-field scattering intensity (Eq.~\ref{eq: scattering intensity}) and the Maxwell stress tensor evaluation (Eq.~\ref{eq: stress tensor average}), are obtained from these via Fourier analysis over one period of the incident wave. The scattered field is obtained by subtracting the incident field from the total field. Further implementation details can be found in \cite{khan2024electromagnetic, hauser2017stable}.

The computational domain is a square of side length $L$ in the $x$--$y$ plane, with the Janus cylinder positioned at its center (Fig.~\ref{fig:schematic_LBM}). A plane wave is introduced through the left boundary, while the remaining boundaries allow outgoing waves to exit the domain. In all simulations, the relative permeability is set to $\mu_r = 1$.

The spatial resolution is chosen based on the dielectric properties of the system. For scattering calculations of the metallo-dielectric Janus cylinder with $\varepsilon_{r_1} = 1, 2$ and $5$, the resolution within the dielectric region is set to $\lambda_{\varepsilon_{r_1}} / \Delta x = 50$, where $\lambda_{\varepsilon_{r_1}}$ is the wavelength inside the dielectric. The domain size is taken as $L/a = 10$. For radiation force and torque calculations, we use $a/\Delta x = 40$ with $L/a = 4$ for metallo-dielectric Janus cylinders and $a/\Delta x = 50$ with $L/a = 4$ for dielectric Janus cylinders. As shown in \cite{khan2024electromagnetic}, maintaining $\lambda/\Delta x \ge 10$ ensures that radiation force calculations remain within $5\%$ of analytical results. Since the maximum dielectric constant considered here is $\varepsilon_{r_2} = 25$ (corresponding to $\varepsilon_{r_1} = 5$ and $\varepsilon_{r_2}/\varepsilon_{r_1} = 5$), the chosen resolution satisfies this criterion.

Higher dielectric contrasts lead to stronger internal reflections, requiring longer simulation times to reach steady state. Convergence is assessed by monitoring the total electromagnetic energy in the computational domain, which approaches a constant value, and by verifying that the computed radiation force and torque settle into a sinusoidal oscillation of constant amplitude at twice the frequency of the incident wave (arising from the quadratic dependence of the Maxwell stress tensor on the fields), consistent with the time-averaged formulation.

The numerical stability of the LBM framework has been demonstrated in our previous work \cite{khan2024electromagnetic}, where simulations were shown to remain stable over long integration times without spurious growth in electromagnetic energy. In the present study, all simulations exhibit stable behavior. Non-reflecting open boundary conditions are implemented using a first-order extrapolation scheme \cite{khan2026lbm_em_scattering}, with residual boundary reflections confirmed to stabilize at levels of $10^{-3}$ or lower and decreasing further with increasing spatial resolution.


\section{Results and discussion} \label{sec: results and discussions}

In this section, we present a detailed analysis of the radiation force and torque on Janus cylinders. We first validate the LBM by comparing the computed far-field scattering intensity with analytical solutions. For the metallo-dielectric Janus cylinder, analytical expressions for radiation drag, lift, and torque are derived and used to benchmark the numerical results. The analysis is then extended to dielectric Janus cylinders to examine the effects of permittivity contrast and interface orientation on the radiation response. To gain physical insight, we analyze two-dimensional maps of the force and torque and relate the observed variations to resonance and the angular redistribution of the scattered field, with supporting slice analysis provided in the appendices. Finally, we investigate the coupled translational and rotational dynamics of Janus cylinders in a viscous fluid at low Reynolds number, using the computed forces and torque to determine particle trajectories.

\subsection{Metallo-dielectric Janus cylinder} \label{sec:MDJC}

\subsubsection{Far-field scattering characteristics}

\begin{figure*}[!htb]
    \centering
    \subfigure[$\phi_0 = 0$]{\includegraphics[width=0.48\textwidth]{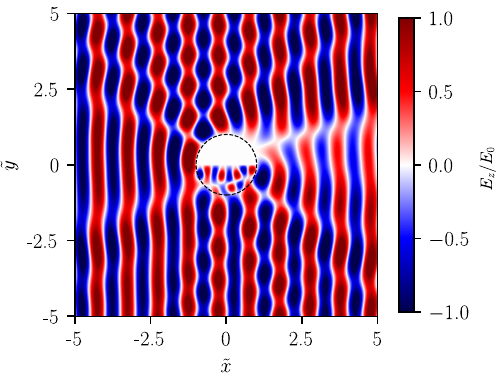}}
    \subfigure[$\phi_0 = - \pi / 4$]{\includegraphics[width=0.48\textwidth]{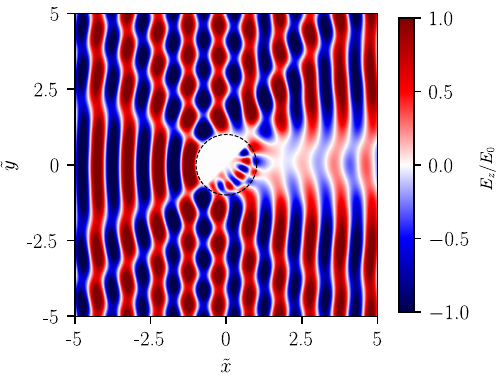}}
    \subfigure[$\varepsilon_{r_1} = 1$]{\includegraphics[width=0.32\textwidth]{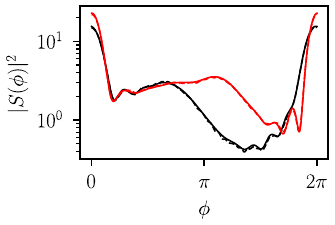}}
    \subfigure[$\varepsilon_{r_1} = 2$]{\includegraphics[width=0.32\textwidth]{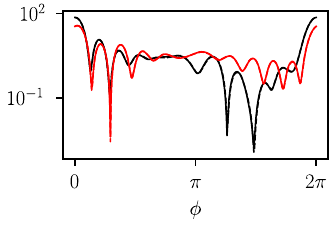}}
    \subfigure[$\varepsilon_{r_1} = 5$]{\includegraphics[width=0.32\textwidth]{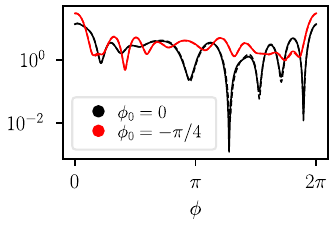}}
    \subfigure[$\varepsilon_{r_1} = 1$]{\includegraphics[width=0.32\textwidth]{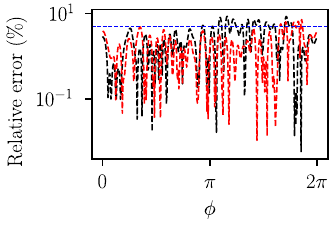}}
    \subfigure[$\varepsilon_{r_1} = 2$]{\includegraphics[width=0.32\textwidth]{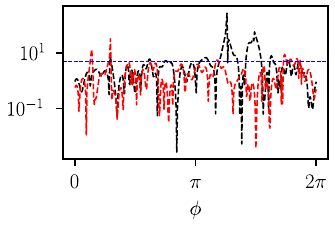}}
    \subfigure[$\varepsilon_{r_1} = 5$]{\includegraphics[width=0.32\textwidth]{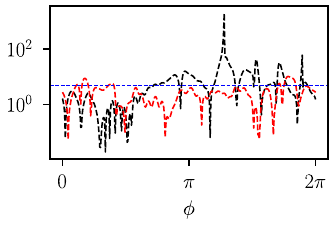}}
    \caption{(Top row) Normalized total electric field ($E_z/E_0$) for a metallo-dielectric Janus cylinder under TM$^z$-polarized plane-wave illumination for two interface orientations: (a) $\phi_0 = 0$ and (b) $\phi_0 = -\pi/4$. The incident wave of wavelength $\lambda$ propagates from the left, and the cylinder radius is $a = \lambda$. The spatial coordinates are normalized as $\tilde{x} = x/a$ and $\tilde{y} = y/a$. (Middle row) Far-field scattering intensity $|S(\phi)|^2$ as a function of the scattering angle $\phi$ for different dielectric constants of the dielectric half: (c) $\varepsilon_{r_1} = 1$, (d) $\varepsilon_{r_1} = 2$, and (e) $\varepsilon_{r_1} = 5$. Black and red curves correspond to $\phi_0 = 0$ and $\phi_0 = -\pi/4$, respectively. Solid lines denote analytical results, while dashed lines represent LBM simulations. (Bottom row) Relative error between LBM and analytical results for the corresponding cases: (f) $\varepsilon_{r_1} = 1$, (g) $\varepsilon_{r_1} = 2$, and (h) $\varepsilon_{r_1} = 5$. The horizontal blue line indicates a $5\%$ error level. Forward and backward scattering correspond to $\phi = 0$ and $\phi = \pi$, respectively.}
    \label{fig:SW}
\end{figure*}

Figures~\ref{fig:SW}(a) and \ref{fig:SW}(b) show the normalized total electric field distributions for a TM$^z$-polarized plane wave incident on a metallo-dielectric Janus cylinder for two interface orientations, $\phi_0 = 0$ and $\phi_0 = -\pi/4$, computed using LBM. In both cases, the dielectric half has $\varepsilon_{r_1} = 5$. The field distributions clearly demonstrate the asymmetry introduced by the Janus configuration, which leads to orientation-dependent scattering patterns.

The corresponding far-field scattering intensity $|S(\phi)|^2$ is shown in Figs.~\ref{fig:SW}(c--e) for three dielectric constants, $\varepsilon_{r_1} = 1, 2,$ and $5$. For each case, results are presented for $\phi_0 = 0$ and $\phi_0 = -\pi/4$. The analytical solutions (solid lines) and LBM results (dashed lines) are in excellent agreement, with the two curves being nearly indistinguishable over most scattering angles. This confirms the ability of the LBM to accurately capture the angular distribution of the scattered field for metallo-dielectric Janus cylinders.

To quantify the agreement, the relative error between the LBM and analytical results is shown in Figs.~\ref{fig:SW}(f–-h). The error is generally below $5\%$ across most scattering angles, as indicated by the horizontal reference line. Higher deviations are observed in localized angular regions where the scattering intensity exhibits sharp variations. These discrepancies arise primarily from the strong angular gradients in the scattered field, where small phase differences between the numerical and analytical solutions can lead to amplified relative errors. Additionally, the present formulation neglects material absorption, which would otherwise smooth these sharp features, and the staircase approximation used to represent the circular cylinder on the Cartesian lattice introduces additional geometric discretization errors. Despite these effects, the overall agreement between LBM and analytical results remains excellent, validating the LBM implementation for far-field scattering calculations on metallo-dielectric Janus cylinders.

For metallo-dielectric Janus cylinders, analytical expressions for the scattered field can be obtained by solving for the coefficients $B_m$ and $D_m$ introduced in Section~\ref{sec:theory}. The total electric field at the cylinder surface ($r = a$), denoted by $E_z^{\text{tot}}$, satisfies $E_z^{\text{tot}} = 0$ over the PEC half. This field is obtained by summing Eqs.~\eqref{eq: incident field} and \eqref{eq: scattered field}, equating the result to $E_z^{\text{tot}}$, and applying orthogonality conditions to yield:
\begin{subequations} \label{eq:constants_Bm_Dm}
\begin{align}
    B_m &= \frac{\alpha_m}{2 \pi H_m^{(1)} (k_0 a) } \int_0^{\pi} E_z^{\text{tot}} (\phi) \cos(m \phi) \, d\phi \notag\\
    &\quad - \frac{\alpha_m (-j)^m J_m (k_0 a) \cos(m \phi_0)}{H_m^{(1)} (k_0 a)}, \label{eq:constant_Bm} \\
    D_m &= \frac{1}{\pi H_m^{(1)} (k_0 a) } \int_0^{\pi} E_z^{\text{tot}} (\phi) \sin(m \phi) \, d\phi \notag\\
    &\quad - \frac{2 (-j)^m J_m (k_0 a) \sin(m \phi_0)}{H_m^{(1)} (k_0 a)}. \label{eq:constant_Dm}
\end{align}
\end{subequations}

The unknown surface field $E_z^{\text{tot}}$ is obtained by solving the integral equation derived from the continuity boundary condition \cite{hurd1975diffraction}:
\begin{equation} \label{eq: integral equation}
    \int_0^{\pi} E_z^{\text{tot}} (\phi') K (\phi, \phi') d \phi' = Y (\phi),
\end{equation}
where the kernel $K (\phi, \phi')$ and forcing term $Y (\phi)$ are given by:
\begin{align}
    K (\phi, \phi') &= \sum_{m = 0}^{\infty} \left( \frac{H^{(1)'}_m(k_0 a)}{H^{(1)}_m(k_0 a)} - 2 \frac{k}{k_0} \frac{J_m^{'} (k a)}{J_m (k a)} \right) \sin(m \phi) \sin(m \phi') \notag\\
    &\quad + \frac{\alpha_m}{2} \frac{H^{(1)'}_m(k_0 a)}{H^{(1)}_m(k_0 a)} \cos(m \phi) \cos(m \phi'), \\
    Y (\phi) &= \frac{2 j}{k_0 a} \sum_{m = 0}^{\infty} \frac{\alpha_m (-j)^m \cos\{m (\phi - \phi_0)\}}{H^{(1)}_m (k_0 a)},
\end{align}
where $k = k_0 \sqrt{\varepsilon_{r_1}}$ is the wavenumber in the dielectric, and primes denote differentiation with respect to the argument. The derivation also involves the internal field within the dielectric region; details can be found in \cite{hurd1975diffraction}.

Discretizing Eq.~\eqref{eq: integral equation} leads to the linear system
\begin{equation} \label{eq: discrete integral equation}
    \sum_{n = 1}^N K_{mn} (E_z^{\text{tot}})_n \, \Delta \phi = Y_m ,
\end{equation}
which can be written in matrix form as
\begin{equation} \label{eq: matrix integral equation}
    \mathbb{K} \, \mathbf{E}_z^{\text{tot}}  = \frac{1}{\Delta \phi} {\bf Y},
\end{equation}
or equivalently:
\begin{equation} \label{eq: solving Etot}
    \mathbf{E}_z^{\text{tot}} = \frac{1}{\Delta \phi} \, \mathbb{K}^{-1} {\bf Y},
\end{equation}
where $\mathbf{E}_z^{\text{tot}}$ and $\mathbf{Y}$ are $N \times 1$ vectors, $\mathbb{K}$ is an $N \times N$ matrix, and $\Delta \phi = \pi/(N-1)$. Solving this system yields the surface field $\mathbf{E}_z^{\text{tot}}$.

The coefficients $B_m$ and $D_m$ are then evaluated using Eq.~\eqref{eq:constants_Bm_Dm} and substituted into Eq.~\eqref{eq: scattered field} to compute the scattered field and the corresponding far-field scattering intensity.

\subsubsection{Radiation drag, lift, and torque}

\begin{figure*}[!htb]
    \centering
    \subfigure[\( \tilde{F}_x^{LBM} \)]{\includegraphics[width=0.35\textwidth]{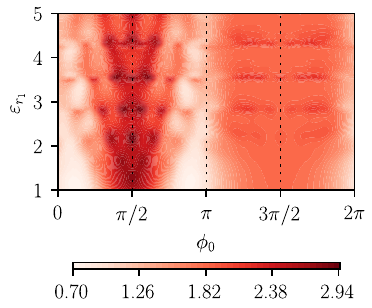}}
    \subfigure[\( \tilde{F}_y^{LBM} \)]{\includegraphics[width=0.31\textwidth]{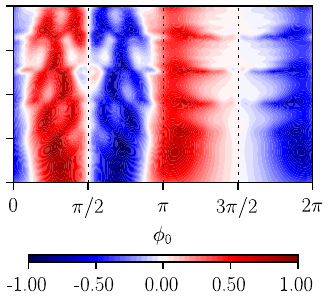}}
    \subfigure[\( \tilde{M}_z^{LBM} \)]{\includegraphics[width=0.31\textwidth]{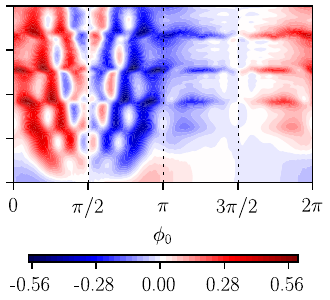}}
    \subfigure[\( \left| \frac{ \tilde{F}_x^{Exact} - \tilde{F}_x^{LBM} }{\tilde{F}_x^{Exact}} \right| \times 100 \)]{\includegraphics[width=0.35\textwidth]{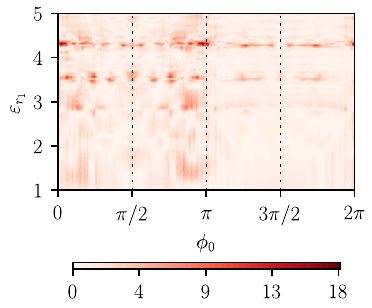}}
    \subfigure[\( \left| \frac{ \tilde{F}_y^{Exact} - \tilde{F}_y^{LBM} }{\tilde{F}_y^{Exact}} \right| \times 100 \)]{\includegraphics[width=0.31\textwidth]{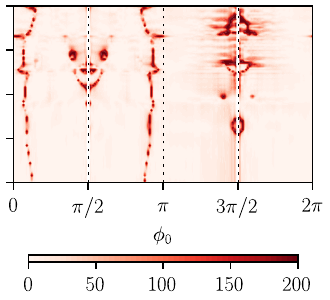}}
    \subfigure[\( \left| \frac{ \tilde{M}_z^{Exact} - \tilde{M}_z^{LBM} }{\tilde{M}_z^{Exact}} \right| \times 100 \)]{\includegraphics[width=0.31\textwidth]{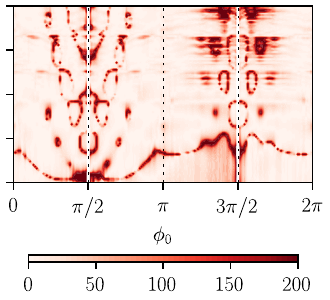}}
    \caption{Top row: Non-dimensional (a) radiation drag, (b) radiation lift, and (c) radiation torque, on a metallo-dielectric Janus cylinder as a function of dielectric constant of the dielectric half ($\varepsilon_{r_1}$) and orientation angle ($\phi_0$) calculated using LBM. Bottom row: Comparison between results from LBM and analytical solutions: Error in (d) radiation drag, (e) radiation lift, and (f) radiation torque.}
    \label{fig:HS75_Map}
\end{figure*}

Figures~\ref{fig:HS75_Map}(a–-c) present the radiation drag $\tilde{F}_x$, lift $\tilde{F}_y$, and torque $\tilde{M}_z$ for a metallo-dielectric Janus cylinder under TM$^z$-polarized plane-wave illumination, computed using LBM. These quantities are shown as functions of the interface orientation $\phi_0$ (horizontal axis), varied from $0$ to $2\pi$ in increments of $\pi/90$, and the dielectric constant $\varepsilon_{r_1}$ (vertical axis), varied from 1 to 5 in steps of 0.04. The contour colours represent magnitude and sign, with red indicating positive values and blue negative values.

Analytical expressions for the radiation force and torque are obtained by substituting the coefficients $B_m$ and $D_m$ into the total field expressions (Eq.~\ref{eq:total_field}) and evaluating the Maxwell stress tensor (Eq.~\ref{eq: stress tensor average}) in the force and torque expressions (Eqs.\ref{eq: drag} -- \ref{eq: Torque}). Although the resulting expressions initially involve double summations due to the quadratic dependence on the fields, the orthogonality of trigonometric functions reduces them to single summations, yielding the compact forms given below.

The force components can be written as
\begin{equation} \label{eq:Fx}
    \tilde{F}_x = \mathcal{U}\cos\phi_0 + \mathcal{V}\sin\phi_0,
\end{equation}
\begin{equation} \label{eq:Fy}
    \tilde{F}_y = -\mathcal{U}\sin\phi_0 + \mathcal{V}\cos\phi_0.
\end{equation}
where $\mathcal{U}$ and $\mathcal{V}$ are given by Eqs.~\eqref{eq:U} and \eqref{eq:V} below, with the coefficients $C_m$, $S_m$, $C_m'$, $S_m'$ defined in Eqs.~\ref{eq:coefficients}.
\begin{widetext}
    \begin{align}
        \mathcal{U} &= \{\Re(C_1C_0^*) - \Re(S_1S_0^*) + \Re(C_1'C_0'^*) - \Re(S_1'S_0'^*)\} +\sum_{m=0}^{\infty}\{\Re(C_mC_{m+1}^*) + 
        \Re(S_mS_{m+1}^*) + \Re(C_m'C_{m+1}'^*) + 
        \Re(S_m'S_{m+1}'^*)\} \notag \\
        &\quad - \frac{1}{(k_0a)^2}\sum_{m=0}^{\infty} 
        m(m+1)\{\Re(C_mC_{m+1}^*) + \Re(S_mS_{m+1}^*)\} \notag \\
        &\quad - \frac{1}{k_0a}\Re\!\left[\{S_0'S_1^* - C_0'C_1^*\} 
        - \sum_{m=1}^{\infty} m\{(C_{m-1}' - C_{m+1}')C_m^* + 
        (S_{m-1}' - S_{m+1}')S_m^*\}\right] \label{eq:U} \\
        \mathcal{V} &= \Re(C_1S_0^*) + \Re(C_0S_1^*) + \Re(C_1'S_0'^*) + \Re(S_1'C_0'^*) + \sum_{m=0}^{\infty}\{\Re(S_{m+1}C_m^*) - 
        \Re(S_mC_{m+1}^*) + \Re(C_m'S_{m+1}'^*) - 
        \Re(S_m'C_{m+1}'^*)\} \notag \\
        &\quad + \frac{1}{(k_0a)^2}\sum_{m=0}^{\infty} 
        m(m+1)\{\Re(C_{m+1}S_m^*) - \Re(C_mS_{m+1}^*)\} \notag \\
        &\quad + \frac{1}{k_0a}\Re\!\left[\{C_0'S_1^* + S_0'C_1^*\} 
        + \sum_{m=1}^{\infty} m\{(C_{m-1}' + C_{m+1}')S_m^* - 
        (S_{m-1}' + S_{m+1}')C_m^*\}\right]
    \label{eq:V}
    \end{align}
\end{widetext}

The radiation torque is given by
\begin{equation} \label{eq:Torque}
    \tilde{M}_z = \frac{1}{k_0 a} \sum_{m=1}^{\infty} 
    m\,\Re\!\left(S_m' C_m^* - C_m' S_m^*\right).
\end{equation}

The normalized forces and torque are defined in Section~\ref{sec:theory}. The coefficients appearing in Eqs.~\eqref{eq:U}--\eqref{eq:Torque} are defined as
\begin{subequations} \label{eq:coefficients}
    \begin{gather}
        C_m = \left[ \alpha_m (-j)^m J_m (k_0 a) \cos{(m \phi_0)} + B_m H_m^{(1)} (k_0 a) \right], \\
        S_m = \left[ \alpha_m (-j)^m J_m (k_0 a) \sin{(m \phi_0)} + D_m H_m^{(1)} (k_0 a) \right], \\
        C_m' = \left[ \alpha_m (-j)^m J_m' (k_0 a) \cos{(m \phi_0)} + B_m H_m^{(1)'} (k_0 a) \right], \\
        S_m' = \left[ \alpha_m (-j)^m J_m' (k_0 a) \sin{(m \phi_0)} + D_m H_m^{(1)'} (k_0 a) \right].
    \end{gather}
\end{subequations}

A comparison between LBM and analytical results is shown in Figs.~\ref{fig:HS75_Map}(d–-f). The percentage error remains small over most of the $(\varepsilon_{r_1}, \phi_0)$ space, with larger deviations occurring in regions of sharp spatial variation or where lift and torque approach zero. These deviations arise from the sensitivity of interference-driven quantities to small phase errors, the absence of material absorption (which would otherwise smooth sharp features), and the staircase approximation of the curved boundary. Overall, the agreement confirms the accuracy of the LBM.

\begin{figure}[!htb]
    {\includegraphics[]{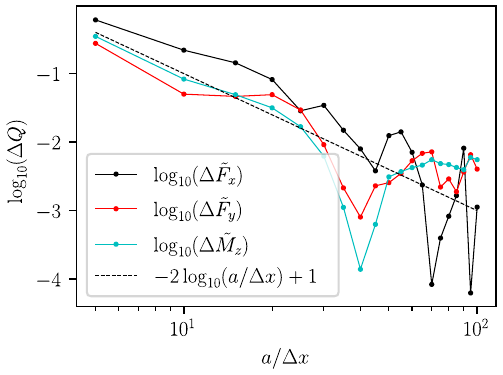}}
    \caption{Grid convergence of the LBM results for radiation forces and torque for a metallo-dielectric Janus cylinder at $\phi_0 = 7\pi/36$ and $\varepsilon_{r_1} = 2$. The logarithm of the errors in drag ($\Delta \tilde{F}_x$), lift ($\Delta \tilde{F}_y$), and torque ($\Delta \tilde{M}_z$) are plotted as functions of grid resolution ($a/\Delta x$) by comparison with analytical solutions, where $\Delta Q = |Q^{\text{LBM}} - Q^{\text{exact}}|/|Q^{\text{exact}}|$ denotes the relative error for $Q \in \{\tilde{F}_x, \tilde{F}_y, \tilde{M}_z\}$. The dashed line indicates second-order convergence. Despite fluctuations at higher resolutions due to multipolar interference and geometric discretization effects, the overall trend confirms near second-order accuracy.}
    \label{fig:convergence}
\end{figure}

The convergence of the computed radiation force and torque with respect to spatial resolution is assessed by comparison with analytical solutions for the metallo-dielectric Janus cylinder at a representative parameter combination of $\phi_0 = 7\pi/36$ and $\varepsilon_{r_1} = 2$, as shown in Fig.~\ref{fig:convergence}. The error decreases approximately with second-order accuracy as the grid is refined, and both the energy convergence and force and torque convergence criteria are satisfied in all simulations presented here.

The drag force $\tilde{F}_x$ is symmetric, while the lift $\tilde{F}_y$ and torque $\tilde{M}_z$ are antisymmetric about $\phi_0 = \pi/2$ and $3\pi/2$, reflecting the symmetry of the geometry and the rotational structure of the force expressions.

The normalized drag varies between approximately $0.7$ and $2.94$, with a maximum when the dielectric half faces the incident wave ($\phi_0 = \pi/2$) and a minimum when the interface is aligned with the incident direction ($\phi_0 = \pi$). When the PEC half faces the wave ($\phi_0 = 3\pi/2$), the drag approaches that of a homogeneous PEC circular cylinder of the same radius $a$.

The lift and torque exhibit strong orientation dependence, with larger variations when the dielectric half faces the incident wave and smoother behavior for PEC-facing configurations. Both quantities change sign near $\phi_0 = \pi/2$ and $3\pi/2$, with sharper transitions near $\pi/2$. These variations reflect the sensitivity of the radiation response to scattering asymmetry and modal interference. Notably, the torque response undergoes a qualitative change as $\varepsilon_{r_1}$ increases: oscillatory features progressively emerge, leading to additional zero-crossings that correspond to new stable and unstable equilibrium orientations. 

It is also worth noting that the maximum values of the normalized drag, lift, and torque for the metallo-dielectric Janus cylinder — approximately $2.94$, $1.00$, and $0.56$ respectively — establish a reference scale for the optomechanical response. As shown in the following section, these maximum values are largely preserved in purely dielectric Janus cylinders across a wide range of material contrasts, suggesting that the magnitude of the radiation response is primarily set by the size-to-wavelength ratio $a/\lambda = 1$ rather than by the specific material composition of the two halves.

\subsection{Dielectric Janus cylinder}

\subsubsection{Radiation drag, lift, and torque}

Following the validation of the LBM for metallo-dielectric Janus cylinders, we now extend the analysis to dielectric Janus cylinders under TM$^z$-polarized plane-wave illumination.

The dielectric Janus cylinder consists of two halves with dielectric constants $\varepsilon_{r_1}$ (region II) and $\varepsilon_{r_2}$ (region III), embedded in vacuum. To examine the effects of dielectric contrast and interface orientation, we consider three representative cases with $\varepsilon_{r_1} = 1$, $2$, and $5$. For each case, the dielectric contrast is varied through the ratio $\varepsilon_{r_2}/\varepsilon_{r_1}$ from $1$ to $5$ in increments of $0.04$, while the interface orientation $\phi_0$ is varied from $0$ to $2\pi$ in steps of $\pi/90$.

\begin{figure*}[!htb]
    \centering
    \subfigure[$\tilde{F}_x$ for fixed $\varepsilon_{r_1} = 1$]{\includegraphics[width=0.35\textwidth]{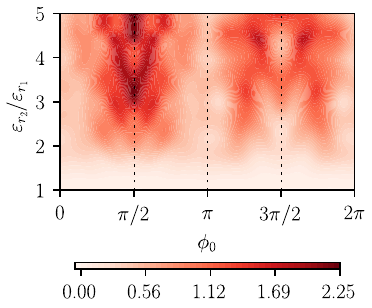}}
    \subfigure[$\tilde{F}_x$ for fixed $\varepsilon_{r_1} = 2$]{\includegraphics[width=0.31\textwidth]{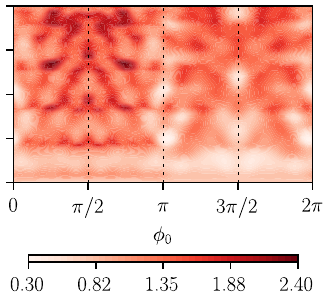}}
    \subfigure[$\tilde{F}_x$ for fixed $\varepsilon_{r_1} = 5$]{\includegraphics[width=0.31\textwidth]{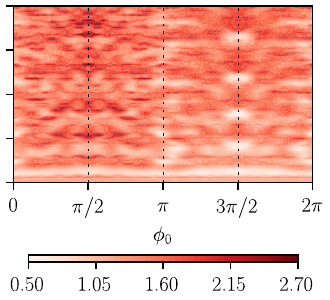}}
    \subfigure[$\tilde{F}_y$ for fixed $\varepsilon_{r_1} = 1$]{\includegraphics[width=0.35\textwidth]{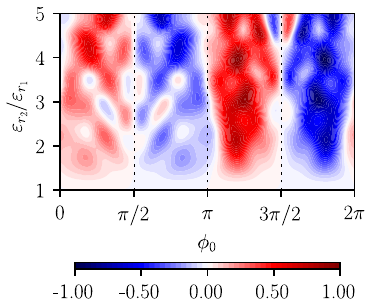}}
    \subfigure[$\tilde{F}_y$ for fixed $\varepsilon_{r_1} = 2$]{\includegraphics[width=0.31\textwidth]{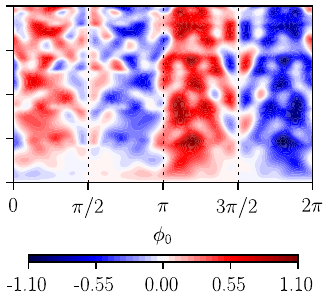}}
    \subfigure[$\tilde{F}_y$ for fixed $\varepsilon_{r_1} = 5$]{\includegraphics[width=0.31\textwidth]{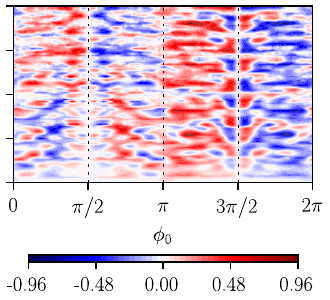}}
    \subfigure[$\tilde{M}_z$ for fixed $\varepsilon_{r_1} = 1$]{\includegraphics[width=0.35\textwidth]{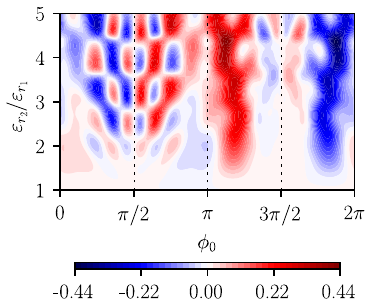}}
    \subfigure[$\tilde{M}_z$ for fixed $\varepsilon_{r_1} = 2$]{\includegraphics[width=0.31\textwidth]{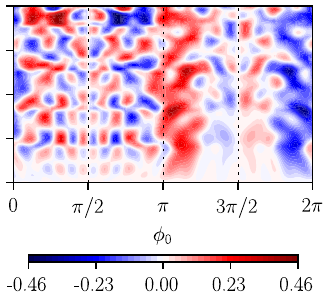}}
    \subfigure[$\tilde{M}_z$ for fixed $\varepsilon_{r_1} = 5$]{\includegraphics[width=0.31\textwidth]{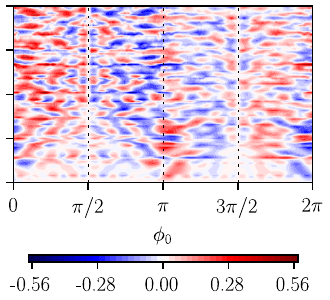}}
    \caption{Non-dimensional radiation drag ($\tilde{F}_x$, top row), lift ($\tilde{F}_y$, middle row), and torque ($\tilde{M}_z$, bottom row) on a dielectric Janus cylinder as functions of the orientation angle $\phi_0$ (horizontal axis, $0$–$2 \pi$) and the dielectric constant ratio $\varepsilon_{r_2} / \varepsilon_{r_1}$ (vertical axis, 1–5), with the dielectric constant of region II ($\varepsilon_{r_1}$) held fixed. Results are shown for three values of $\varepsilon_{r_1}$: left column (a, d, g) $\varepsilon_{r_1} = 1$, middle column (b, e, h) $\varepsilon_{r_1} = 2$, and right column (c, f, i) $\varepsilon_{r_1} = 5$.}
    \label{fig:2D_map}
\end{figure*}

Figure~\ref{fig:2D_map} shows the non-dimensional radiation drag (top row), lift (middle row), and torque (bottom row) as functions of the dielectric contrast (vertical axis) and orientation angle (horizontal axis). The left, middle, and right columns correspond to $\varepsilon_{r_1} = 1$, $2$, and $5$, respectively. For $\varepsilon_{r_1} = 1$, the configuration reduces to a semi-circular dielectric cylinder, where region II matches the surrounding medium.

The drag force $\tilde{F}_x$ remains symmetric, while the lift $\tilde{F}_y$ and torque $\tilde{M}_z$ are antisymmetric about $\phi_0 = \pi/2$ and $3\pi/2$, consistent with the symmetry observed in the metallo-dielectric case. As $\varepsilon_{r_1}$ increases, the force and torque distributions develop progressively sharper features and finer structures.

Quantitatively, for $\varepsilon_{r_1} = 1$ [Fig.~\ref{fig:2D_map}(a)], the drag varies from $0$ to approximately $2.25$, with the minimum occurring at $\varepsilon_{r_2}/\varepsilon_{r_1} = 1$, where scattering vanishes due to uniform material properties. For $\varepsilon_{r_1} = 2$ [Fig.~\ref{fig:2D_map}(b)], the drag ranges from about $0.3$ to $2.4$, while for $\varepsilon_{r_1} = 5$ [Fig.~\ref{fig:2D_map}(c)], it varies between approximately $0.5$ and $2.7$. Although the maximum drag remains comparable across all three cases --- ranging from approximately $2.25$ to $2.70$ --- and is also consistent with the maximum drag observed for the metallo-dielectric Janus cylinder, the minimum drag increases with $\varepsilon_{r_1}$, leading to a reduced dynamic range. Similarly, the peak lift magnitude remains in the range $0.96$--$1.10$ and the peak torque magnitude in the range $0.44$--$0.56$ across the three cases, consistent with the maximum values of approximately $1.00$ and $0.56$ respectively observed for the metallo-dielectric Janus cylinder. The peak magnitudes of all three quantities therefore remain of order unity across all configurations, indicating that the overall scale of the optomechanical response is governed primarily by the size-to-wavelength ratio rather than by the specific material contrast or the presence of a metallic half. What changes substantially with increasing $\varepsilon_{r_1}$ is not the magnitude but the structural complexity of the response --- sharper gradients, finer features, and more localized extrema emerge as the dielectric contrast increases, reflecting the progressive contribution of higher interference orders to the scattered field redistribution.

The force and torque maps presented in Figs.~\ref{fig:HS75_Map} and \ref{fig:2D_map} provide a comprehensive representation of the optomechanical response of Janus cylinders as functions of interface orientation and dielectric contrast. These maps effectively serve as design diagrams, indicating the expected magnitude and direction of radiation drag, lift, and torque for a given material configuration and orientation. Such information can be useful in guiding experimental efforts, where selecting appropriate material contrasts and particle orientations is critical for achieving desired force and torque characteristics. It should be noted that all maps are computed at a fixed size-to-wavelength ratio $a/\lambda = 1$; varying the wavelength would shift the dominant multipolar contributions and modify the quantitative features of the maps, although the underlying sensitivity to interface orientation and dielectric contrast is expected to persist across the intermediate scattering regime. In this sense, the present results offer a systematic framework for understanding and predicting radiation-driven behavior in asymmetric particles under idealized conditions.

\subsubsection{Physical mechanisms governing force and torque variations}

To elucidate the sharp features and large differences between the local maxima and minima observed in the contour maps, we examine representative scattered field distributions for two cases: (1) varying the orientation angle $\phi_0$ while fixing all other parameters, and (2) varying the dielectric constant ratio $\varepsilon_{r_2}/\varepsilon_{r_1}$ while fixing all other parameters.

\begin{figure*}[!htb]
    \centering
    \subfigure[$\varepsilon_{r_2} / \varepsilon_{r_1} = 4$, $\phi_0 = \pi / 180$.]{\includegraphics[width=0.48\textwidth]{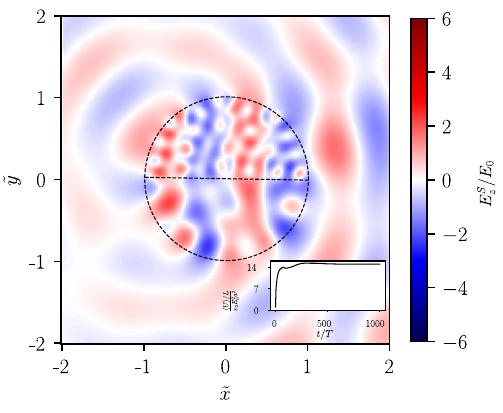}}
    \subfigure[$\varepsilon_{r_2} / \varepsilon_{r_1} = 4$, $\phi_0 = 11 \pi / 30$.]{\includegraphics[width=0.48\textwidth]{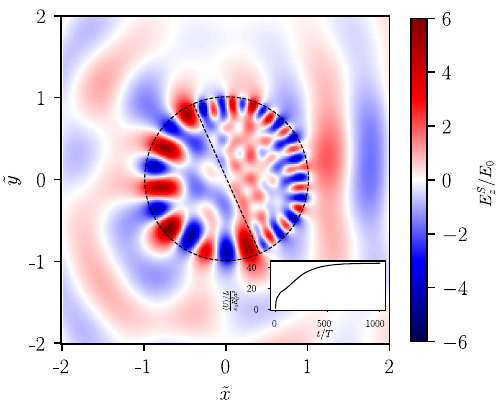}}
    \subfigure[$\varepsilon_{r_2} / \varepsilon_{r_1} = 3.60$, $\phi_0 =\pi / 2$.]{\includegraphics[width=0.48\textwidth]{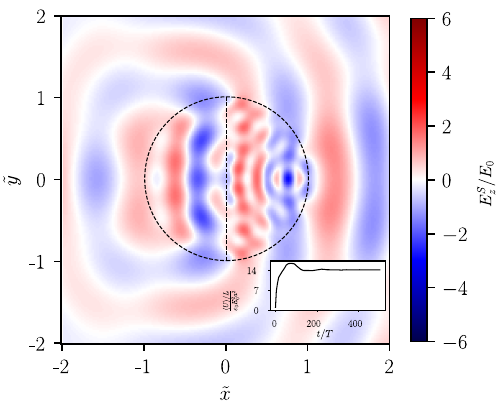}}
    \subfigure[$\varepsilon_{r_2} / \varepsilon_{r_1} = 3.66$, $\phi_0 =\pi / 2$.]{\includegraphics[width=0.48\textwidth]{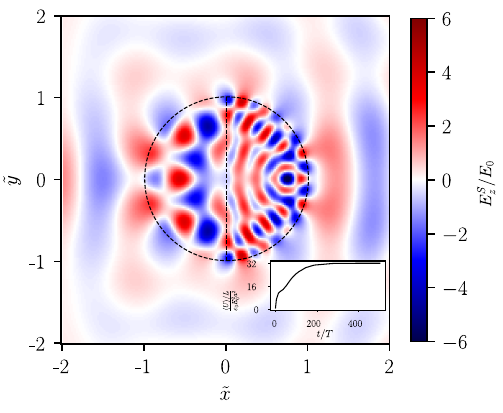}}
    \caption{Normalized scattered electric field ($E_z^S/E_0$) for a dielectric Janus cylinder under TM$^z$-polarized plane-wave illumination, with the dielectric constant of region~II fixed at $\varepsilon_{r_1} = 5$. Top row: dielectric constant ratio $\varepsilon_{r_2} / \varepsilon_{r_1} = 4$ for two different interface orientations: (a) $\phi_0 = \pi / 180$ and (b) $\phi_0 = 11\pi / 30$. Bottom row: interface orientation fixed at $\phi_0 = \pi / 2$ for two different dielectric constant ratios: (c) $\varepsilon_{r_2} / \varepsilon_{r_1} = 3.60$ and (d) $\varepsilon_{r_2} / \varepsilon_{r_1} = 3.66$. In all cases, the incident plane wave of wavelength $\lambda$ propagates from left to right, and the cylinder radius is $a = \lambda$. Insets display the normalized total energy in the computational domain versus time, illustrating convergence to the steady state.}
    \label{fig:DJ_peak}
\end{figure*}

For Case 1, we set $\varepsilon_{r_1} = 5$ and $\varepsilon_{r_2}/\varepsilon_{r_1} = 4$, identifying the minimum and maximum drag at $\phi_0 = \pi/180$ and $\phi_0 = 11\pi/30$, with values of approximately $0.8$ and $1.73$ respectively. The corresponding scattered electric fields and total domain energy are shown in Figs.~\ref{fig:DJ_peak}(a,b). At $\phi_0 = 11\pi/30$, both the amplitude of the scattered electric field and the total energy in the domain are substantially higher than at $\phi_0 = \pi/180$ (approximately $14$ at $\phi_0 = \pi/180$ versus $40$ at $\phi_0 = 11\pi/30$), consistent with resonance-driven energy amplification where specific interface orientations trigger enhanced internal reflections and greater energy storage within the dielectric region.

For Case 2, we fix $\varepsilon_{r_1} = 5$ and $\phi_0 = \pi/2$, and compare two nearby dielectric contrast values: $\varepsilon_{r_2}/\varepsilon_{r_1} = 3.60$ (drag valley) and $3.66$ (drag peak), shown in Figs.~\ref{fig:DJ_peak}(c,d). A similar trend is observed, where the case with maximum drag shows both a higher amplitude in the scattered electric field and greater total energy in the domain (approximately $14$ at $\varepsilon_{r_2}/\varepsilon_{r_1} = 3.60$ versus $32$ at $\varepsilon_{r_2}/\varepsilon_{r_1} = 3.66$), again consistent with resonance-driven energy amplification occurring at specific values of dielectric contrast.

These observations demonstrate that resonance can be triggered by varying either the interface orientation or the dielectric contrast, resulting in amplified scattered fields and increased radiation forces and torque. However, large differences in force and torque cannot always be attributed to resonance alone; they may also arise from changes in the distribution of the scattered field around the Janus cylinder, as illustrated in Appendix~\ref{appendix b} for the metallo-dielectric case.

\subsection{Trajectories of Janus cylinders}

\begin{figure*}[!htb]
    \centering
    \subfigure[$\varepsilon_{r_1} = 1$]{\includegraphics[width=0.33\textwidth]{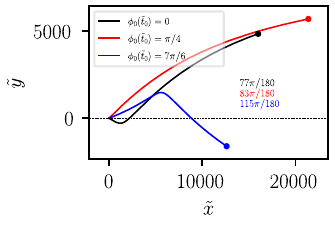}}
    \subfigure[$\varepsilon_{r_1} = 2$]{\includegraphics[width=0.33\textwidth]{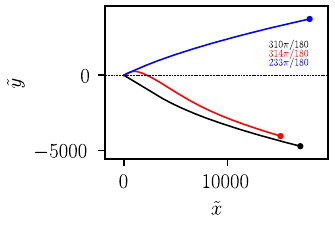}}
    \subfigure[$\varepsilon_{r_1} = 5$]{\includegraphics[width=0.33\textwidth]{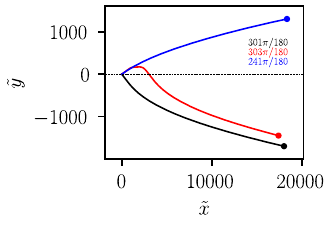}}
    \subfigure[$\varepsilon_{r_1} = 1$]{\includegraphics[width=0.33\textwidth]{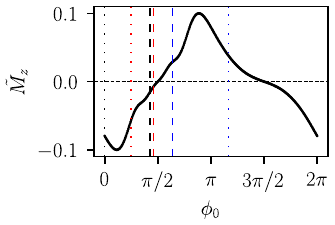}}
    \subfigure[$\varepsilon_{r_1} = 2$]{\includegraphics[width=0.33\textwidth]{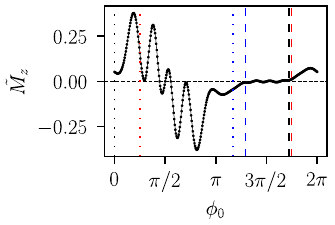}}
    \subfigure[$\varepsilon_{r_1} = 5$]{\includegraphics[width=0.33\textwidth]{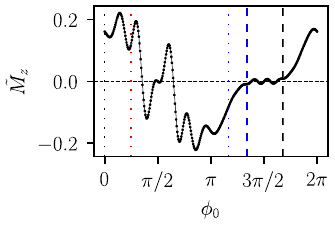}}
    \caption{Trajectories of the metallo-dielectric Janus cylinder in the $\tilde{x}$--$\tilde{y}$ plane for three initial interface orientations. The dielectric constant of the dielectric half is (a) $\varepsilon_{r_1}=1$, (b) $\varepsilon_{r_1}=2$, and (c) $\varepsilon_{r_1}=5$. Markers indicate the final positions, and the corresponding interface orientations at $\tilde{t}=\tilde{t}_f$ are labeled. (d--f) Corresponding radiation torque $\tilde{M}_z$ as a function of the interface orientation $\phi_0$ for $\varepsilon_{r_1}=1, 2,$ and $5$, respectively. Vertical dotted lines denote the initial orientations, while dashed lines indicate the final orientations reached at $\tilde{t}=\tilde{t}_f$.}

    \label{fig:MDJC_trajectory}
\end{figure*}

\begin{figure*}[!htb]
    \centering
    \subfigure[$\varepsilon_{r_1} = 1, \varepsilon_{r_2} / \varepsilon_{r_1} = 2$]{\includegraphics[width=0.33\textwidth]{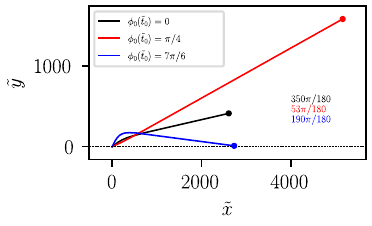}}
    \subfigure[$\varepsilon_{r_1} = 2, \varepsilon_{r_2} / \varepsilon_{r_1} = 2$]{\includegraphics[width=0.33\textwidth]{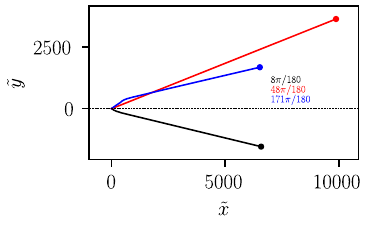}}
    \subfigure[$\varepsilon_{r_1} = 5, \varepsilon_{r_2} / \varepsilon_{r_1} = 2$]{\includegraphics[width=0.33\textwidth]{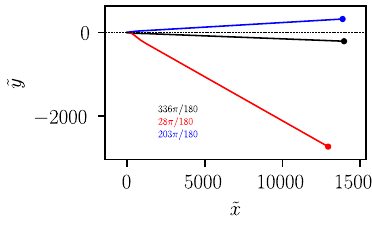}}
    \subfigure[$\varepsilon_{r_1} = 1, \varepsilon_{r_2} / \varepsilon_{r_1} = 2$]{\includegraphics[width=0.33\textwidth]{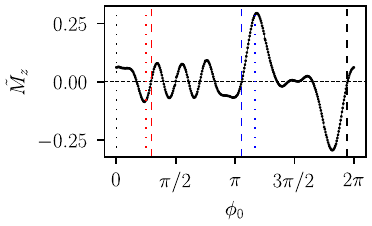}}
    \subfigure[$\varepsilon_{r_1} = 2, \varepsilon_{r_2} / \varepsilon_{r_1} = 2$]{\includegraphics[width=0.33\textwidth]{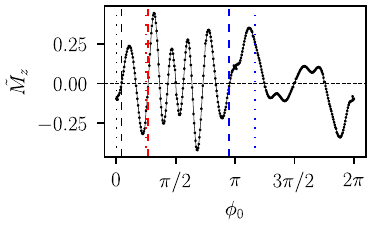}}
    \subfigure[$\varepsilon_{r_1} = 5, \varepsilon_{r_2} / \varepsilon_{r_1} = 2$]{\includegraphics[width=0.33\textwidth]{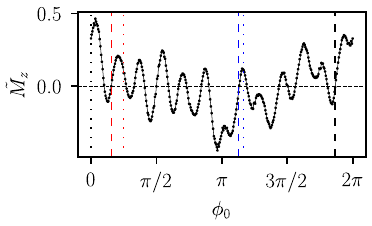}}
    \caption{Trajectories of the dielectric Janus cylinder in the $\tilde{x}$--$\tilde{y}$ plane for three initial interface orientations with the dielectric contrast ratio of the two halves fixed at $\varepsilon_{r_2}/\varepsilon_{r_1}=2$. The dielectric constant of one half is (a) $\varepsilon_{r_1}=1$, (b) $\varepsilon_{r_1}=2$, and (c) $\varepsilon_{r_1}=5$. Markers indicate the final positions, and the corresponding interface orientations at $\tilde{t}=\tilde{t}_f$ are labeled. (d--f) Corresponding radiation torque $\tilde{M}_z$ as a function of the interface orientation $\phi_0$ for $\varepsilon_{r_1}=1, 2,$ and $5$, respectively. Vertical dotted lines denote the initial orientations, while dashed lines indicate the final orientations reached at $\tilde{t}=\tilde{t}_f$.}

    \label{fig:DJC_trajectory}
\end{figure*}

To investigate the motion of Janus cylinders under electromagnetic radiation, we consider their translation and rotation in a viscous fluid. In the low Reynolds number regime, inertial effects are negligible. Since the particle velocity is much smaller than the speed of light, the Minkowski magnetoelectric coupling term (proportional to $v/c$) is neglected \cite{gordon1973radiation}. The instantaneous force and torque balances are given by
\begin{equation} \label{eq:force-torque-balance}
\mathbf{F}^H + \mathbf{F}^{EM} = 0, \qquad
\mathbf{M}^H + \mathbf{M}^{EM} = 0,
\end{equation}
where $\mathbf{F}^{EM}$ and $\mathbf{M}^{EM}$ denote radiation force and torque, and $\mathbf{F}^H$ and $\mathbf{M}^H$ represent hydrodynamic drag and viscous torque.

To regularize the Stokes paradox for an infinitely long cylinder, we adopt a slender-body approximation for a finite cylinder of length $l \gg a$. The hydrodynamic resistance is then given by \cite{guazzelli2011physical}
\begin{equation} \label{eq:hydro-force-torque}
\mathbf{F}^H = - \frac{4 \pi \eta \mathbf{U}}{\log(l/a)}, \qquad
\mathbf{M}^H = - 4 \pi \eta a^2 \boldsymbol{\Omega},
\end{equation}
where $\eta$ is the fluid viscosity, $\mathbf{U}$ is the translational velocity, and $\boldsymbol{\Omega}$ is the angular velocity.

Substituting Eq.~\eqref{eq:hydro-force-torque} into Eq.~\eqref{eq:force-torque-balance} yields
\begin{equation}
U_x = \frac{\log(l/a)\,\tilde{F}_x F_0}{4 \pi \eta}, \quad
U_y = \frac{\log(l/a)\,\tilde{F}_y F_0}{4 \pi \eta}, \quad
\Omega_z = \frac{\tilde{M}_z M_0}{4 \pi \eta a^2},
\end{equation}
where $U_x$ and $U_y$ are the translational velocities in the $x$ and $y$ directions respectively, $\Omega_z$ is the angular velocity, $\tilde{F}_x$ and $\tilde{F}_y$ are the normalized radiation drag and lift, $\tilde{M}_z$ is the normalized radiation torque, and $F_0$ and $M_0$ are the reference scales defined in Section~\ref{sec:theory}.

Using the non-dimensional variables $(\tilde{x},\tilde{y})=(x/a,y/a)$ and $\tilde{t}=t/T_0$ with $T_0=a\eta/F_0$, the kinematic equations become
\begin{equation}
\frac{d\tilde{x}}{d\tilde{t}} = \frac{\log(l/a)}{4\pi}\tilde{F}_x, 
\quad
\frac{d\tilde{y}}{d\tilde{t}} = \frac{\log(l/a)}{4\pi}\tilde{F}_y,
\quad
\frac{d\phi_0}{d\tilde{t}} = \frac{\tilde{M}_z}{2\pi}.
\end{equation}

The trajectories are obtained by integrating these equations using the force and torque fields computed from LBM or analytical solutions. All trajectories are initialized at $(\tilde{x}, \tilde{y}) = (0,0)$ with prescribed initial orientations $\phi_0(\tilde{t}_0)$ and integrated up to $\tilde{t}_f = 10,000$.

The force and torque maps governing the dynamics are shown in Figs.~\ref{fig:HS75_Map} and \ref{fig:2D_map} for metallo-dielectric and dielectric Janus cylinders, respectively. Trajectories are computed for representative parameter values. For the metallo-dielectric case, $\varepsilon_{r_1}=1,2,$ and $5$ are considered, while for the dielectric Janus cylinder the ratio is fixed at $\varepsilon_{r_2}/\varepsilon_{r_1}=2$ with the same set of $\varepsilon_{r_1}$. The initial orientations are chosen as $\phi_0(\tilde{t}_0) = 0, \pi/4,$ and $7\pi/6$, deliberately avoiding the symmetry configurations $\phi_0 = \pi/2$ and $3\pi/2$ where the torque vanishes by symmetry and no reorientation occurs; the selected orientations therefore ensure non-zero initial torque and non-trivial reorientation dynamics throughout the simulation.

The resulting trajectories are shown in Figs.~\ref{fig:MDJC_trajectory} and \ref{fig:DJC_trajectory}. In each case, the top row shows motion in the $\tilde{x}$--$\tilde{y}$ plane, while the bottom row shows the corresponding torque $\tilde{M}_z(\phi_0)$. The dotted and dashed lines indicate the initial ($\phi_0(\tilde{t}_0)$) and final ($\phi_0(\tilde{t}_f)$) orientations, respectively.

The trajectory evolution is governed by the torque landscape. The sign of $\tilde{M}_z$ determines the direction of rotation, while its magnitude controls the rate of reorientation. The interface angle evolves toward torque-free configurations ($\tilde{M}_z = 0$), which act as attractors in the orientational dynamics. However, depending on the torque magnitude and the finite simulation time $\tilde{t}_f$, the particle may not fully reach these equilibria — in particular, for the metallo-dielectric case where the torque varies smoothly with $\phi_0$, the reorientation is gradual and the final orientation at $\tilde{t}_f$ may not coincide exactly with a torque-free state.

The curvature of the trajectory arises from the coupling between orientation and force components. During reorientation, $\tilde{M}_z$ drives the change in interface orientation $\phi_0$, which continuously modifies both $\tilde{F}_x$ and $\tilde{F}_y$, producing curved paths. Once the torque becomes small, the orientation stabilizes and the motion transitions to nearly linear translation governed by the force evaluated at the attained orientation. Since $\tilde{F}_x$ remains positive, forward motion is always maintained.

A clear distinction is observed between metallo-dielectric and dielectric Janus cylinders. In the metallo-dielectric case, the torque varies smoothly with $\phi_0$, exhibiting comparatively few zero-crossings, leading to gradual reorientation and sustained curvature. In contrast, dielectric Janus cylinders exhibit a significantly larger number of zero-crossings, which increases progressively with $\varepsilon_{r_1}$, resulting in rapid alignment with the nearest equilibrium orientation and shorter curved segments followed by nearly straight trajectories. This orientation-dependent trajectory shaping is a direct consequence of the broken symmetry of the Janus particle: the asymmetric scattering generates non-zero torque and lateral force components that are entirely absent in homogeneous cylinders, where symmetry suppresses both quantities under plane-wave illumination. The trajectories presented here should therefore be interpreted as illustrative consequences of the computed force and torque landscapes rather than as evidence of robust controllability, given the idealized nature of the hydrodynamic model employed.

The present trajectory analysis is based on a simplified hydrodynamic model and neglects fluctuations, Brownian motion, and perturbations in orientation and position. In addition, material absorption and thermal effects are not included, and the metallic region is modeled as a perfect electric conductor. The results therefore represent scattering-dominated dynamics under idealized conditions.

\subsection{Modeling assumptions and limitations}

It is worth noting that many experimental Janus systems --- particularly hybrid plasmonic particles such as those studied in \cite{gonzalezcolsa2022nanojet, serrera20253} --- are dominated by absorption and thermoplasmonic effects, where asymmetric heating drives self-thermophoretic motion rather than scattering-induced momentum transfer. The present results therefore isolate and quantify the scattering contribution, which provides a complementary perspective to absorption-dominated systems and is relevant to purely dielectric Janus particles where thermoplasmonic effects are absent. The parameter regime considered is compatible with experimentally accessible optical manipulation setups: for micron-scale particles illuminated at visible to near-infrared wavelengths, typical optical intensities of $\sim 10^6$--$10^8$ W/m$^2$ used in optical tweezers can generate measurable radiation forces and torques \cite{zhou2022recent}, and dielectric Janus particles with controlled material contrast can be fabricated using established techniques such as vapor deposition or colloidal synthesis \cite{zhang2017janus, xiao2018review}.

While the present model assumes lossless dielectrics and a PEC limit for the metallic half, real materials introduce dispersion and absorption that would modify the quantitative force and torque values and may introduce additional thermoplasmonic contributions. The PEC approximation is appropriate for highly conducting metals such as gold or silver at near-infrared wavelengths, where the skin depth is much smaller than the cylinder radius and the reflectivity approaches unity \cite{maier2007plasmonics}. However, for smaller particles or at shorter wavelengths, material dispersion and ohmic losses become significant, leading to absorption-driven contributions to the radiation force that are not captured by the present model. In such cases, the scattering-only results presented here provide a lower bound on the total optomechanical response, with absorption-driven contributions adding to the net force and torque. Furthermore, all results are obtained at a fixed size-to-wavelength ratio $a/\lambda = 1$; decreasing $a/\lambda$ toward the Rayleigh regime would suppress higher-order multipole contributions and smooth the force and torque maps, while increasing it toward the geometric optics regime would recover ray-tracing behavior \cite{hulst1981light, bohren1998absorption}, suggesting that the intermediate regime studied here produces the richest and most orientation-sensitive optomechanical response.


\section{Conclusion} \label{sec: Conclusion}

We have investigated the radiation drag, lift, and torque on metallo-dielectric and dielectric Janus cylinders under TM$^z$-polarized plane-wave illumination at $a/\lambda = 1$, where full-wave solutions of Maxwell's equations are required. The lattice Boltzmann method (LBM) was employed and validated against analytical solutions for the metallo-dielectric case, showing excellent agreement in far-field scattering intensity, radiation force, and torque across a wide range of dielectric constants and interface orientations.

For the metallo-dielectric Janus cylinder, the drag peaks when the dielectric half faces the incoming wave, while drag, lift, and torque exhibit stronger fluctuations in this configuration and smoother variations when the PEC half is exposed. Furthermore, when the PEC side faces the incident wave, the drag recovers that of a homogeneous PEC cylinder of the same radius. These trends demonstrate the critical role of interface orientation and dielectric constant in shaping the radiation response of metallo-dielectric systems. Notably, as the dielectric constant increases, the torque response undergoes a continuous qualitative transition in which oscillatory features progressively emerge, giving rise to additional torque-free orientations that represent new stable and unstable equilibria.

For the dielectric Janus cylinder, we fixed the dielectric constant of one half ($\varepsilon_{r_1}$) and varied both the dielectric contrast ($\varepsilon_{r_2}/\varepsilon_{r_1}$) and the orientation $\phi_0$. Across the three cases considered ($\varepsilon_{r_1}=1,2,5$), the peak magnitudes of drag, lift, and torque remain comparable to each other and are also consistent with those observed for the metallo-dielectric Janus cylinder, indicating that the overall scale of the optomechanical response is governed primarily by the size-to-wavelength ratio $a/\lambda = 1$ rather than by the specific material composition. What changes substantially with increasing $\varepsilon_{r_1}$ is not the magnitude but the structural complexity of the response --- sharper gradients, finer features, and more localized extrema emerge as the dielectric contrast increases, reflecting the progressive contribution of higher interference orders to the scattered field distribution. Beyond characterizing the optomechanical response, these force and torque maps serve directly as design diagrams: given a material configuration and interface orientation, the maps immediately predict the expected drag, lift, torque, and consequently the particle trajectory, providing a systematic tool for guiding the selection of material parameters and orientations in experimental realizations.

Another key outcome of this work is the identification of two distinct physical mechanisms governing force and torque variations in both metallo-dielectric and dielectric Janus cylinders: resonance-driven enhancement, associated with constructive interference and increased energy storage within the dielectric region, and scattered field redistribution, where changes in the angular distribution of the scattered field modify the net momentum transfer independently of the total stored energy. These two mechanisms are demonstrated in Appendix~\ref{appendix b} through representative scattered field distributions and total stored energy, which together provide a physically transparent distinction between the two cases. Such resonances, by enhancing the total stored energy within the dielectric region, provide potential routes for increasing specific force components through appropriate choice of material parameters and interface orientation.

Finally, we coupled the computed electromagnetic forces and torques with hydrodynamic resistance in the Stokes regime to examine the trajectories of Janus cylinders in viscous fluids. The results reveal diverse dynamical behaviors, including curved paths and rapid reorientation in regimes where torque is significant, and nearly straight-line motion when the torque approaches zero. Dielectric contrast influences both the translational velocity and the curvature of trajectories, with torque sign changes leading to complex path patterns. These results demonstrate how radiation-induced forces and torque give rise to directed motion and trajectory shaping of Janus particles under plane-wave illumination. The broken symmetry of the Janus particle is essential to these behaviors: the asymmetric scattering generates non-zero torque and lateral force components that are entirely absent in homogeneous cylinders under plane-wave illumination. We emphasize that the present system represents an externally actuated, non-equilibrium system rather than an active particle in the strict sense, as the motion is driven by externally applied electromagnetic fields without local energy transduction. The presented trajectories should be interpreted as illustrative consequences of the computed force and torque landscapes under idealized conditions, rather than as evidence of robust controllability. These findings provide physical insight into scattering-driven optomechanical responses of Janus particles and suggest potential applications in optofluidic transport and directed assembly.


\section*{Conflicts of interest}

There are no conflicts to declare.

\section*{Data availability}

The data that support the findings of this study are available within the article. The code is available at \url{https://github.com/mohd-meraj-khan}.



\appendix


\section{Slice Analysis of Force and Torque Contours for Metallo-dielectric Janus Cylinders}\label{appendix a}

\begin{figure*}[!htb]
    \centering
    \subfigure[Radiation drag $\tilde{F}_x$ vs. orientation angle $\phi_0$ for fixed $\varepsilon_{r_1}$.]{\includegraphics[width=0.33\textwidth]{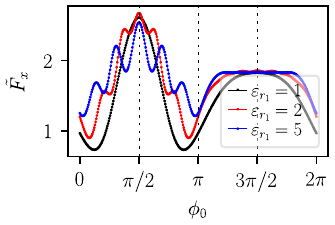}}
    \subfigure[Radiation lift $\tilde{F}_y$ vs. orientation angle $\phi_0$ for fixed $\varepsilon_{r_1}$.]{\includegraphics[width=0.33\textwidth]{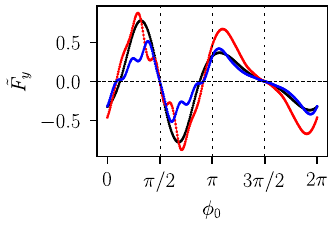}}
    \subfigure[Radiation torque $\tilde{M}_z$ vs. orientation angle $\phi_0$ for fixed $\varepsilon_{r_1}$.]{\includegraphics[width=0.33\textwidth]{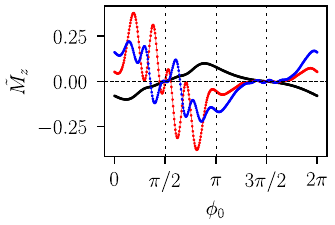}}
    \subfigure[Radiation drag $\tilde{F}_x$ vs. dielectric constant $\varepsilon_{r_1}$ for fixed $\phi_0$.]{\includegraphics[width=0.33\textwidth]{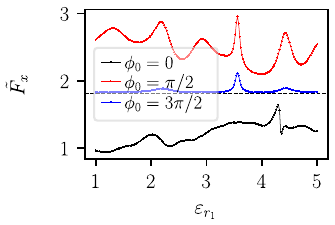}}
    \subfigure[Radiation lift $\tilde{F}_y$ vs. dielectric constant $\varepsilon_{r_1}$ for fixed $\phi_0$.]{\includegraphics[width=0.33\textwidth]{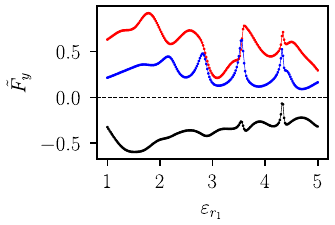}}
    \subfigure[Radiation torque $\tilde{M}_z$ vs. dielectric constant $\varepsilon_{r_1}$ for fixed $\phi_0$.]{\includegraphics[width=0.33\textwidth]{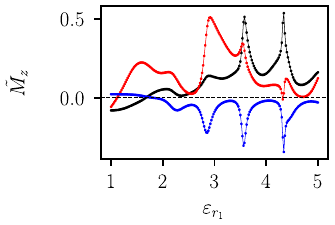}}
    \caption{Horizontal (top row) and vertical (bottom row) slices of the non-dimensional radiation drag ($\tilde{F}_x$, left column), lift ($\tilde{F}_y$, middle column), and torque ($\tilde{M}_z$, right column) on a metallo-dielectric Janus cylinder, extracted from Figs.~\ref{fig:HS75_Map}(a--c). The top row (a–-c) shows the dependence on orientation angle $\phi_0$ for fixed dielectric constants $\varepsilon_{r_1} = 1$ (black), $2$ (red), and $5$ (blue). The bottom row (d–-f) shows the dependence on dielectric constant $\varepsilon_{r_1}$ for fixed orientations $\phi_0 =0$ (black), $\pi / 2$ (red), and $3 \pi / 2$ (blue).}
    \label{fig:MDJC_slice}
\end{figure*}

To provide additional insight, Fig.~\ref{fig:MDJC_slice} shows horizontal and vertical slices of the radiation drag, lift, and torque contour plots for the metallo-dielectric Janus cylinder, generated with a finer resolution of $\Delta \phi_0 = \pi/180$ and $\Delta \varepsilon_{r_1} = 0.02$ compared to $\Delta \phi_0 = \pi/90$ and $\Delta \varepsilon_{r_1} = 0.04$ used in the contour plots of the main text. The horizontal slices (top row) show the dependence on orientation angle $\phi_0$ at fixed $\varepsilon_{r_1}$, while the vertical slices (bottom row) show the dependence on $\varepsilon_{r_1}$ at fixed $\phi_0$.

For the horizontal slices, results are shown for $\varepsilon_{r_1} = 1$, $2$, and $5$. When $\varepsilon_{r_1} = 1$, the dielectric half has the same dielectric constant as the surrounding medium, reducing the system to a semi-circular PEC cylinder. The corresponding drag, lift, and torque (black lines) vary smoothly across all orientations. The drag is symmetric while the lift and torque are antisymmetric about $\phi_0 = \pi/2$ and $3\pi/2$, and all three quantities exhibit more fluctuations in the range $0 \leq \phi_0 \leq \pi$ and are smoother in the range $\pi \leq \phi_0 \leq 2\pi$ for $\varepsilon_{r_1} > 1$, reflecting the asymmetric role of the two halves under illumination.

For the vertical slices, results are shown for three fixed orientations: $\phi_0 = 0$ (interface aligned with the incident wave, PEC half oriented along $+y$), $\phi_0 = \pi/2$ (dielectric half facing the incident wave), and $\phi_0 = 3\pi/2$ (PEC half facing the incident wave). As shown in Figs.~\ref{fig:MDJC_slice}(a) and \ref{fig:MDJC_slice}(d), when the PEC half faces the incident wave, the drag on the metallo-dielectric Janus cylinder closely matches that of a homogeneous circular PEC cylinder of radius $a$ (indicated by the horizontal dashed line in Fig.~\ref{fig:MDJC_slice}(d)). The bottom row further shows that the drag, lift, and torque all exhibit sharp fluctuations as $\varepsilon_{r_1}$ is varied at fixed orientation angles, consistent with the resonance and field redistribution mechanisms identified in Appendix~\ref{appendix b}.

\begin{figure*}[!htb]
    \centering
    \subfigure[Radiation drag $\tilde{F}_x$ vs. orientation angle $\phi_0$ for fixed $\varepsilon_{r_1}$.]{\includegraphics[width=0.33\textwidth]{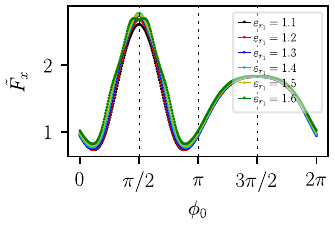}}
    \subfigure[Radiation lift $\tilde{F}_y$ vs. orientation angle $\phi_0$ for fixed $\varepsilon_{r_1}$.]{\includegraphics[width=0.33\textwidth]{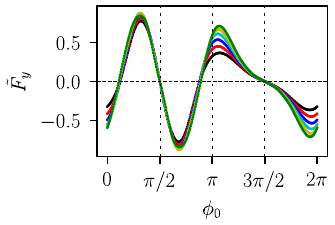}}
    \subfigure[Radiation torque $\tilde{M}_z$ vs. orientation angle $\phi_0$ for fixed $\varepsilon_{r_1}$.]{\includegraphics[width=0.33\textwidth]{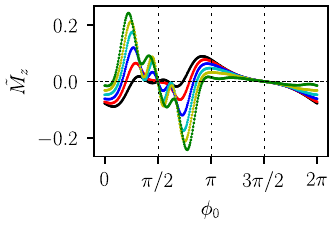}}
    \caption{Same as the top row of Fig.~\ref{fig:MDJC_slice}, showing the angular dependence of the radiation drag ($\tilde{F}_x$), lift ($\tilde{F}_y$), and torque ($\tilde{M}_z$) for a metallo-dielectric Janus cylinder, but for dielectric constants in the range $1 \le \varepsilon_{r_1} \le 2$. The plots highlight the progressive emergence of oscillatory features in the torque with increasing $\varepsilon_{r_1}$, leading to additional extrema and zero-crossings that indicate new equilibrium orientations. In contrast, the drag and lift components remain comparatively smooth, emphasizing that the transition is primarily governed by changes in angular momentum transfer.}

    \label{fig:MDJC_slice_1-2}
\end{figure*}

To further elucidate the qualitative change in the optomechanical response with increasing dielectric contrast, Fig.~\ref{fig:MDJC_slice_1-2} shows angular slices of $\tilde{F}_x$, $\tilde{F}_y$, and $\tilde{M}_z$ for $\varepsilon_{r_1}$ in the range $1 \leq \varepsilon_{r_1} \leq 2$. For $\varepsilon_{r_1} \approx 1$, all three components vary smoothly with $\phi_0$, closely resembling the response of a semi-circular PEC cylinder. As $\varepsilon_{r_1}$ increases, oscillatory features progressively emerge in all three quantities, indicating a continuous transition in the optomechanical response. A key distinction arises in the torque behavior: the oscillations in $\tilde{M}_z$ lead to the appearance of additional local extrema and zero-crossings corresponding to new torque-free orientations, where each new zero-crossing represents the birth of a new pair of stable and unstable equilibrium orientations. In contrast, the drag and lift components remain comparatively smooth throughout this transition, confirming that the qualitative change in reorientation dynamics is governed primarily by changes in angular momentum transfer rather than by overall energy amplification.

\section{Scattered Field Distributions as the Origin of Force Fluctuations in Metallo-dielectric Janus Cylinders}\label{appendix b}

\begin{figure*}[!htb]
    \centering
    \subfigure[ $\varepsilon_{r_1} = 2$, $\phi_0 = 7 \pi / 90$.]{\includegraphics[width=0.48\textwidth]{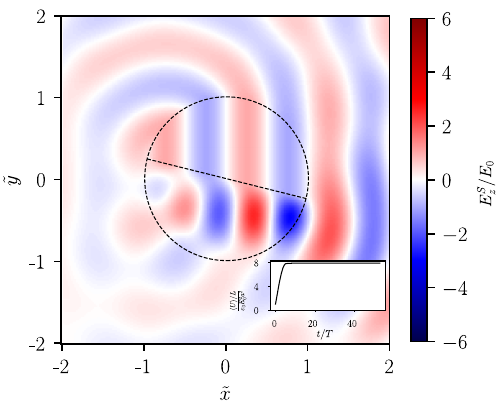}}
    \subfigure[ $\varepsilon_{r_1} = 2$, $\phi_0 =\pi / 2$.]{\includegraphics[width=0.48\textwidth]{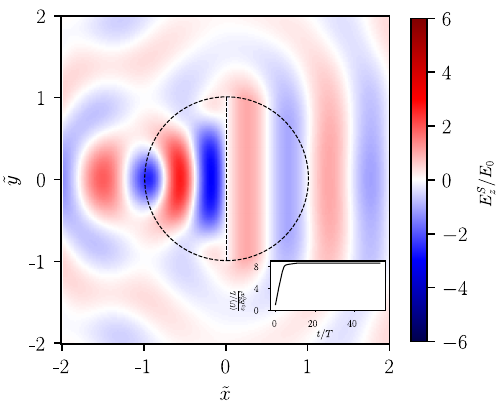}}
    \subfigure[ $\varepsilon_{r_1} = 3.56$, $\phi_0 =\pi / 2$.]{\includegraphics[width=0.48\textwidth]{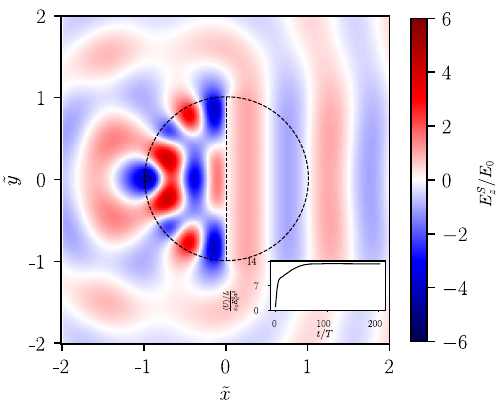}}
    \subfigure[$\varepsilon_{r_1} = 4$, $\phi_0 =\pi / 2$.]{\includegraphics[width=0.48\textwidth]{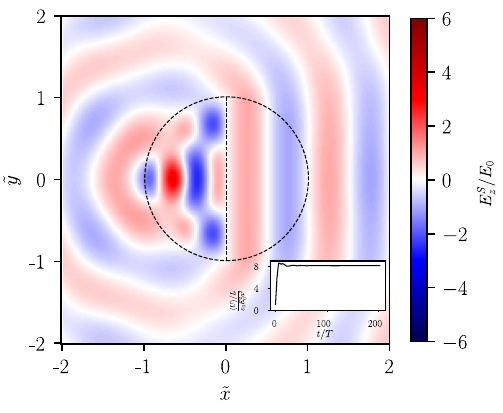}}
    \caption{Normalized scattered electric field ($E_z^S/E_0$) from a metallo-dielectric Janus cylinder under TM$^z$-polarized plane-wave illumination. The top row shows results with the dielectric constant of the dielectric half fixed at $\varepsilon_{r_1} = 2$ while varying the interface orientation: (a) $\phi_0 = 7 \pi / 90$ and (b) $\phi_0 =\pi / 2$. The bottom row shows results with the interface orientation fixed at $\phi_0 =\pi / 2$ while varying the dielectric constant: (c) $\varepsilon_{r_1} = 3.56$ and (d) $\varepsilon_{r_1} = 4$. The inset in each figure displays the normalized total energy in the computational domain as a function of time, illustrating the approach to steady state.}
    \label{fig:MDJC_peak}
\end{figure*}

In Section~\ref{sec: results and discussions}, resonance-driven energy amplification was identified as the mechanism governing force and torque variations in dielectric Janus cylinders, supported by representative scattered field distributions and total stored energy (Fig.~\ref{fig:DJ_peak}). However, large differences in force and torque cannot always be attributed to resonance alone; they may also arise from changes in the spatial distribution of the scattered field even when the total stored energy remains comparable. We now demonstrate both mechanisms for the metallo-dielectric Janus cylinder, considering two cases: (1) fixing all parameters while varying the orientation angle $\phi_0$, and (2) fixing all parameters while varying the dielectric constant $\varepsilon_{r_1}$.

For Case 1, consider Fig.~\ref{fig:MDJC_slice}(a). At $\varepsilon_{r_1} = 2$, the minimum drag of approximately $0.9$ occurs at $\phi_0 = 7\pi/90$ and the maximum drag of approximately $2.68$ occurs at $\phi_0 = \pi/2$, a nearly threefold difference. The corresponding scattered electric fields and total normalized energy are shown in Figs.~\ref{fig:MDJC_peak}(a) and \ref{fig:MDJC_peak}(b). Despite the large difference in drag, the total energy in the domain is nearly identical for both orientations (approximately $8$ in both cases), while the scattered field distributions differ significantly in their spatial structure and angular asymmetry. This confirms that the drag variation here is governed entirely by \textbf{scattered field redistribution}: changes in how the scattered momentum is distributed angularly, rather than any change in the total stored energy.

For Case 2, consider Fig.~\ref{fig:MDJC_slice}(d). At $\phi_0 = \pi/2$, the maximum drag of approximately $2.96$ occurs at $\varepsilon_{r_1} = 3.56$ and the minimum drag of approximately $2.1$ is observed at $\varepsilon_{r_1} = 4$. The corresponding scattered electric fields and total normalized energy are shown in Figs.~\ref{fig:MDJC_peak}(c) and \ref{fig:MDJC_peak}(d). In contrast to Case 1, the total energy at $\varepsilon_{r_1} = 3.56$ (approximately $14$) is nearly twice that at $\varepsilon_{r_1} = 4$ (approximately $8$), accompanied by a stronger scattered field amplitude. This is consistent with \textbf{resonance-driven energy amplification} \cite{hulst1981light, bohren1998absorption}, where enhanced internal reflections at specific values of dielectric constant cause the dielectric half of the cylinder to trap substantially more energy, leading directly to the increased drag.

These two cases together provide clean and unambiguous demonstrations of the two distinct mechanisms. In Case~1, comparable total energies but markedly different scattered field distributions confirm that scattered field redistribution alone can produce large force variations. In Case~2, the substantially increased stored energy confirms resonance-driven energy amplification as the dominant mechanism. The same two mechanisms therefore govern force fluctuations in both dielectric and metallo-dielectric Janus cylinders, and can be distinguished directly by jointly examining the total stored energy and the spatial structure of the scattered field.

\section{Slice Analysis of Force and Torque Contours for Dielectric Janus Cylinders}\label{appendix c}

\begin{figure*}[!htb]
    \centering
    \subfigure[$\varepsilon_{r_1} = 1$]{\includegraphics[width=0.35\textwidth]{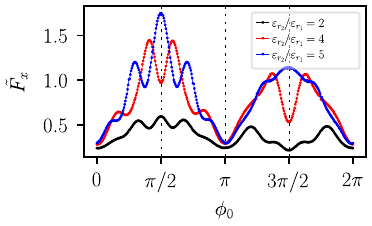}}
    \subfigure[$\varepsilon_{r_1} = 2$]{\includegraphics[width=0.31\textwidth]{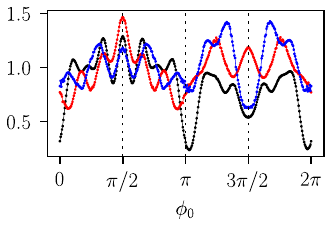}}
    \subfigure[$\varepsilon_{r_1} = 5$]{\includegraphics[width=0.31\textwidth]{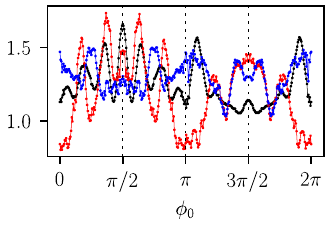}}
    \subfigure[$\varepsilon_{r_1} = 1$]{\includegraphics[width=0.35\textwidth]{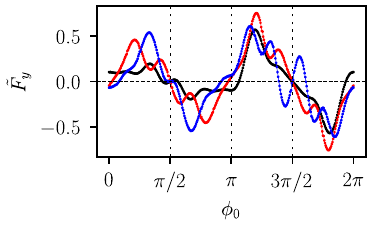}}
    \subfigure[$\varepsilon_{r_1} = 2$]{\includegraphics[width=0.31\textwidth]{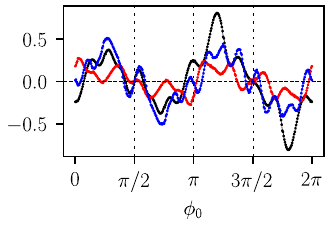}}
    \subfigure[$\varepsilon_{r_1} = 5$]{\includegraphics[width=0.31\textwidth]{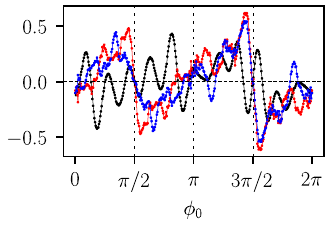}}
    \subfigure[$\varepsilon_{r_1} = 1$]{\includegraphics[width=0.35\textwidth]{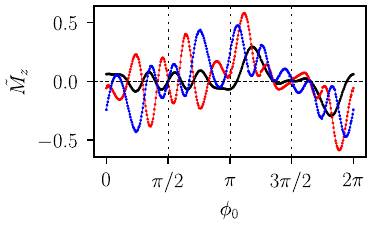}}
    \subfigure[$\varepsilon_{r_1} = 2$]{\includegraphics[width=0.31\textwidth]{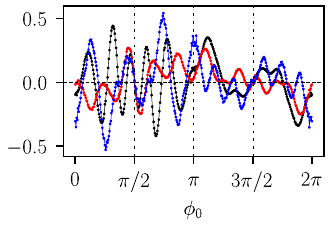}}
    \subfigure[$\varepsilon_{r_1} = 5$]{\includegraphics[width=0.31\textwidth]{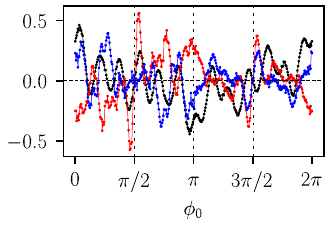}}
    \caption{Non-dimensional radiation drag ($\tilde{F}_x$, top row), lift ($\tilde{F}_y$, middle row), and torque ($\tilde{M}_z$, bottom row) on a dielectric Janus cylinder as functions of the orientation angle $\phi_0$ ($0 - 2 \pi$), extracted as horizontal slices from Fig.~\ref{fig:2D_map}. The dielectric constant of region II ($\varepsilon_{r_1}$) is held fixed with values: (a, d, g) $\varepsilon_{r_1} = 1$, (b, e, h) $\varepsilon_{r_1} = 2$, and (c, f, i) $\varepsilon_{r_1} = 5$, while the dielectric contrast $\varepsilon_{r_2}/\varepsilon_{r_1}$ is varied and shown for values 2 (black), 4 (red), and 5 (blue).}
    \label{fig:DJC_H_slice}
\end{figure*}

\begin{figure*}[!htb]
    \centering
    \subfigure[$\varepsilon_{r_1} = 1$]{\includegraphics[width=0.35\textwidth]{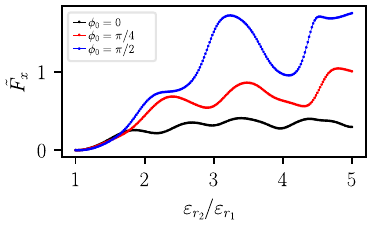}}
    \subfigure[$\varepsilon_{r_1} = 2$]{\includegraphics[width=0.31\textwidth]{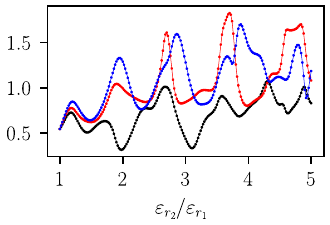}}
    \subfigure[$\varepsilon_{r_1} = 5$]{\includegraphics[width=0.31\textwidth]{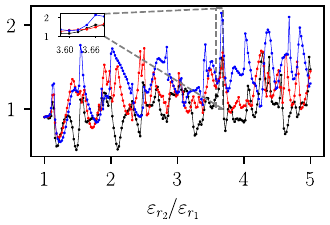}}
    \subfigure[$\varepsilon_{r_1} = 1$]{\includegraphics[width=0.35\textwidth]{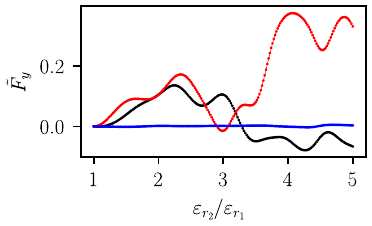}}
    \subfigure[$\varepsilon_{r_1} = 2$]{\includegraphics[width=0.31\textwidth]{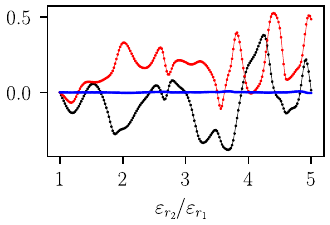}}
    \subfigure[$\varepsilon_{r_1} = 5$]{\includegraphics[width=0.31\textwidth]{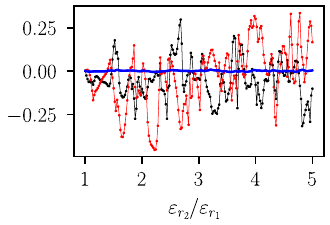}}
    \subfigure[$\varepsilon_{r_1} = 1$]{\includegraphics[width=0.35\textwidth]{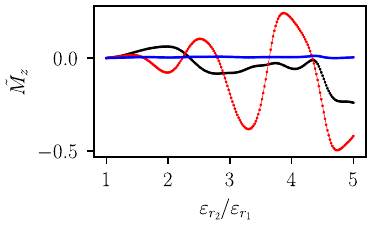}}
    \subfigure[$\varepsilon_{r_1} = 2$]{\includegraphics[width=0.31\textwidth]{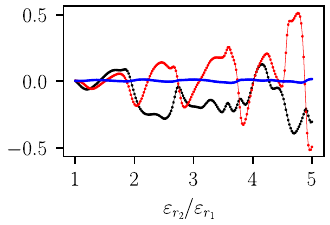}}
    \subfigure[$\varepsilon_{r_1} = 5$]{\includegraphics[width=0.31\textwidth]{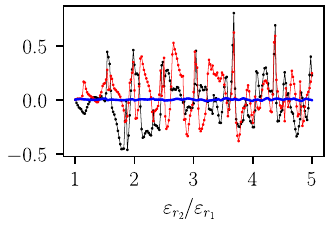}}
    \caption{Non-dimensional radiation drag ($\tilde{F}_x$, top row), lift ($\tilde{F}_y$, middle row), and torque ($\tilde{M}_z$, bottom row) on a dielectric Janus cylinder as functions of the dielectric constant ratio $\varepsilon_{r_2}/\varepsilon_{r_1}$ (1–5), extracted as vertical slices from Fig.~\ref{fig:2D_map}. The orientation angle $\phi_0$ is held fixed at $0$ (black), $\pi / 4$ (red), and $\pi / 2$ (blue). The dielectric constant of region II ($\varepsilon_{r_1}$) is fixed for each column: (a, d, g) $\varepsilon_{r_1} = 1$, (b, e, h) $\varepsilon_{r_1} = 2$, and (c, f, i) $\varepsilon_{r_1} = 5$.}
    \label{fig:DJC_V_slice}
\end{figure*}

Following the approach of Appendix~\ref{appendix a}, we present horizontal and vertical slices of the force and torque contour plots for the dielectric Janus cylinder shown in Fig.~\ref{fig:2D_map}, generated with a finer resolution of $\Delta \phi_0 = \pi/180$ and $\Delta(\varepsilon_{r_2}/\varepsilon_{r_1}) = 0.02$ compared to $\Delta \phi_0 = \pi/90$ and $\Delta(\varepsilon_{r_2}/\varepsilon_{r_1}) = 0.04$ used in the contour plots of the main text.

Figure~\ref{fig:DJC_H_slice} presents horizontal slices of the radiation drag (top row), lift (middle row), and torque (bottom row) as functions of the orientation angle $\phi_0$ for fixed values of the dielectric constant ratio $\varepsilon_{r_2}/\varepsilon_{r_1} = 2$, $4$, and $5$. Each column corresponds to a fixed value of $\varepsilon_{r_1}$: $1$ (left), $2$ (middle), and $5$ (right). In the left column, where $\varepsilon_{r_1} = 1$, region II matches the surrounding medium and the configuration reduces to a semi-circular dielectric cylinder of dielectric constant $\varepsilon_{r_2}$. The drag is symmetric while the lift and torque are antisymmetric about $\phi_0 = \pi/2$ and $3\pi/2$, consistent with the symmetry of the Janus interface. As $\varepsilon_{r_1}$ increases from left to right, the profiles develop increasingly sharp variations and fluctuations while the peak magnitudes remain of comparable order, confirming that it is the structural complexity — not the overall scale — of the response that grows with increasing dielectric contrast, consistent with the observations in the main text.

Figure~\ref{fig:DJC_V_slice} presents vertical slices of the radiation drag (top row), lift (middle row), and torque (bottom row) as functions of the dielectric constant ratio $\varepsilon_{r_2}/\varepsilon_{r_1}$ for fixed orientation angles $\phi_0 = 0$, $\pi/4$, and $\pi/2$, with the same column arrangement as Fig.~\ref{fig:DJC_H_slice}. The profiles become increasingly sharp and fluctuating as $\varepsilon_{r_1}$ increases, particularly evident when moving from the left to the right column. Notably, for a fixed orientation angle, varying the dielectric contrast ratio not only changes the magnitude of the lift and torque but can also produce sign reversals, highlighting the strong sensitivity of these quantities to dielectric contrast. This behavior persists even in the limiting case of the semi-circular dielectric cylinder (left column, $\varepsilon_{r_1} = 1$), further demonstrating that the rich orientation- and contrast-dependent structure of the force and torque response is an intrinsic feature of the intermediate scattering regime rather than a consequence of the two-material Janus geometry alone.

Taken together, the horizontal and vertical slice analyses confirm that the force and torque maps encode information that cannot be inferred from individual parameter sweeps alone, and directly support the design diagram interpretation introduced in the main text.


\bibliography{main}

\end{document}